%% file: main.tex
\documentclass[letterpaper,twocolumn,10pt]{article}
\usepackage{usenix2019_v3}

\usepackage{graphicx}
\usepackage{subcaption}
\usepackage{rotating}
\usepackage{latexsym}
\usepackage{amssymb}
\usepackage{bbm}
\usepackage{pifont}
\usepackage[nointegrals]{wasysym}
\usepackage{stmaryrd}
\usepackage{xcolor}
\usepackage{xurl}
\usepackage{stfloats}
\usepackage{mathtools,stackengine,amsthm}
\usepackage{fontawesome}

\DeclareRobustCommand{\perthousand}{%
  \ifmmode
    \text{\textperthousand}%
  \else
    \textperthousand
  \fi}

\input{prestuff.tex}

\begin{document}

\newcommand{\kg}{{KG}\xspace}
\newcommand{\kgs}{{KGs}\xspace}
\newcommand{\gnns}{{GNNs}\xspace}

\newcommand{\qa}{{\sc QA}\xspace}
\newcommand{\cve}{{CVE}\xspace}
\newcommand{\cves}{{CVEs}\xspace}
\newcommand{\rec}{{\sc Rec}\xspace}
\newcommand{\krl}{{KGR}\xspace}
\newcommand{\kgr}{{KGR}\xspace}
\newcommand{\mrr}{{MRR}\xspace}
\newcommand{\hit}{{HIT@$K$}\xspace}
\newcommand{\hito}{{HIT@$1$}\xspace}
\newcommand{\hitf}{{HIT@$5$}\xspace}

\newcommand{\ndcg}{{NDCG@$K$}\xspace}
\newcommand{\ndcgf}{{NDCG@5}\xspace}

\newcommand{\akp}{{\sc ROAR}$_\mathrm{kp}$\xspace}
\newcommand{\aqp}{{\sc ROAR}$_\mathrm{qm}$\xspace}
\newcommand{\aco}{{\sc ROAR}$_\mathrm{co}$\xspace}
\newcommand{\basone}{{\sc BL}$_\mathrm{1}$\xspace}
\newcommand{\bastwo}{{\sc BL}$_\mathrm{2}$\xspace}

\newcommand{\system}{{\sc ROAR}\xspace}

\newcommand{\NA}{----}
\newcommand{\kgp}{{\sc Kgp}\xspace} 
\newcommand{\lqe}{{\sc Lqe}\xspace} 
\newcommand{\cop}{{\sc CoP}\xspace} 
\newcommand{\ota}{{OTA}\xspace} 
\newcommand{\vect}[1]{\boldsymbol{#1}} 
\newcommand{\sxrightarrow}[2][]{%
  \mathrel{\text{\scriptsize $\xrightarrow[#1]{#2}$}}%
}

\newcommand{\tfont}[1]{$\textsf{\small #1}$}
\newcommand{\minitab}[2][l]{\begin{tabular}{#1}#2\end{tabular}}

\stackMath
\newcommand\xxrightarrow[2][]{\scriptsize \mathrel{%
  \setbox2=\hbox{\stackon{\scriptstyle#1}{\scriptstyle#2}}%
  \stackunder[0pt]{%
    \xrightarrow{\makebox[\dimexpr\wd2\relax]{$\scriptstyle#2$}}%
  }{%
   \scriptstyle#1\,%
  }%
}}

\newcommand\blfootnote[1]{%
  \begingroup
  \renewcommand\thefootnote{}\footnote{#1}%
  \addtocounter{footnote}{-1}%
  \endgroup
}

\theoremstyle{remark}
\newtheorem{example}{Example}

\newcommand{\ting}[1]{\textcolor{black}{#1}\xspace}

\definecolor{zxcolor}{rgb}{0.9, 0.17, 0.31}
\newcommand{\zhaohan}[1]{\textcolor{black}{{#1}}\xspace}
\newcommand{\zhaohantwo}[1]{\textcolor{black}{{#1}}\xspace}
\newcommand{\tianyu}[1]{\textcolor{black}{{#1}}\xspace}

\makeatletter
\newcommand\footnoteref[1]{\protected@xdef\@thefnmark{\ref{#1}}\@footnotemark}
\makeatother

\date{}

\title{\Large \bf On the Security Risks of Knowledge Graph Reasoning}

\author{
{\rm Zhaohan Xi}\\
Penn State
\and
{\rm Tianyu Du}\\
Penn State
\and
{\rm Changjiang Li}\\
Penn State
\and
{\rm Ren Pang}\\
Penn State
\and
{\rm Shouling Ji}\\
Zhejiang University
\and
{\rm Xiapu Luo}\\
Hong Kong Polytechnic University
\and
{\rm Xusheng Xiao}\\
Arizona State University
\and
{\rm Fenglong Ma}\\
Penn State
\and
{\rm Ting Wang}\\
Penn State
} 


\maketitle

\input{abstract.tex}
\input{introduction.tex}
\input{background.tex}

\input{threatmodel.tex}
\input{attack.tex}

\input{experiment.tex}
\input{discussion.tex}
\input{literature.tex}

\input{conclusion.tex}
\clearpage

\bibliographystyle{plain}
\bibliography{cite.bib}

\input{appendix}

\end{document}

%% file: prestuff.tex
\usepackage{epsfig,amsmath,amsfonts,epsfig,multirow,makecell,caption,soul,csquotes,color,wrapfig,subcaption,mathtools,bm,spverbatim,booktabs,tcolorbox,diagbox}
\usepackage[e]{esvect}

\def\tablename{Table}
\def\figurename{Figure}

\usepackage{caption}
\makeatletter
\newif\if@restonecol
\makeatother

\usepackage[boxed, ruled, vlined, linesnumbered]{algorithm2e}
\SetKwRepeat{Do}{do}{while}

\newenvironment{changemargin}[2]{\begin{list}{}{
	\setlength{\topsep}{0pt}\setlength{\leftmargin}{0pt}
	\setlength{\rightmargin}{0pt}
	\setlength{\listparindent}{\parindent}
	\setlength{\itemindent}{\parindent}
	\setlength{\parsep}{0pt plus 1pt}
	\addtolength{\leftmargin}{#1}\addtolength{\rightmargin}{#2}
	}\item}
	{\end{list}}

\usepackage[first=0,last=9]{lcg}
\usepackage{colortbl}
\definecolor{Gray}{gray}{0.8}
\colorlet{Red}{red!10!white}
\colorlet{Blue}{blue!10!white}

\usepackage{hyperref}

\newcommand{\msec}[1]{\S\,\ref{#1}}
\newcommand{\mref}[1]{~\ref{#1}}
\newcommand{\meq}[1]{Eq.\,\ref{#1}}
\newcommand{\mcite}[1]{\cite{#1}}

\newcommand{\meg}{\textit{e.g.}\xspace}
\newcommand{\mie}{\textit{i.e.}\xspace}
\newcommand{\mcf}{\textit{cf}.\xspace}
\newcommand{\mct}[1]{({\it #1})}

\newtcolorbox{mtbox}[1]{left=0.25mm, right=0.25mm, top=0.25mm, bottom=0.25mm, sharp corners, colframe=red!50!black, boxrule=0.5pt, title={#1}, fonttitle=\bfseries, coltitle=red!50!black, attach title to upper={\ --\ }}

\usepackage{scalerel}[2016/12/29]

\makeatletter
\providecommand{\leadsfrom}{%
  \mathrel{\mathpalette\reflect@squig\relax}%
}
\newcommand{\reflect@squig}[2]{%
  \reflectbox{$\m@th#1\leadsto$}%
}
\makeatother

\def\eqref#1{equation~\ref{#1}}

\def\1{\bm{1}}

\DeclareMathAlphabet{\mathsfit}{\encodingdefault}{\sfdefault}{m}{sl}
\SetMathAlphabet{\mathsfit}{bold}{\encodingdefault}{\sfdefault}{bx}{n}

\def\gA{{\mathcal{A}}}

\def\gE{{\mathcal{E}}}

\def\gG{{\mathcal{G}}}

\def\gL{{\mathcal{L}}}

\def\gN{{\mathcal{N}}}

\def\gQ{{\mathcal{Q}}}
\def\gR{{\mathcal{R}}}

\def\gT{{\mathcal{T}}}

\def\gV{{\mathcal{V}}}



%% file: abstract.tex
\begin{abstract}
Knowledge graph reasoning (\krl) -- answering complex logical queries over large knowledge graphs -- represents an important artificial intelligence task, entailing a range of applications (\meg, cyber threat hunting). However, despite its surging popularity, the potential security risks of \krl are largely unexplored, which is concerning, given the increasing use of such capability in security-critical domains.
 
This work represents a solid initial step towards bridging the striking gap. We systematize the security threats to \krl according to the adversary's objectives, knowledge, and attack vectors. Further, we present \system, a new class of attacks that instantiate a variety of such threats. Through empirical evaluation in  representative use cases (\meg, medical decision support, cyber threat hunting, and commonsense reasoning), we demonstrate that \system is highly effective to mislead \krl to suggest pre-defined answers for target queries, yet with negligible impact on non-target ones. Finally, we explore potential countermeasures against \system, including filtering of potentially poisoning knowledge and training with adversarially augmented queries, which leads to several promising research directions.
\end{abstract}

%% file: introduction.tex
\section{Introduction}
\label{sec:intro}

Knowledge graphs (\kgs) are structured representations of human knowledge, capturing real-world objects, relations, and their properties. Thanks to automated KG building tools\mcite{kg-tool}, recent years have witnessed a significant growth of \kgs in various domains (\meg, MITRE\mcite{mitre-attack}, GNBR\mcite{gnbr}, and DrugBank\mcite{drugbank}). One major use of such \kgs is {\em knowledge graph reasoning} (\krl), which answers complex logical queries over \kgs, entailing a range of \zhaohan{applications\mcite{gartner-report} such as information retrieval\mcite{gkg-api}, cyber-threat hunting\mcite{cyscale}, biomedical research\mcite{med-kg-embedding}, and clinical decision support\mcite{qiagen}. 
For instance, \kg-assisted threat hunting has been used in both research prototypes\mcite{apt-detect-sp-2019, end-detect-sp-2020} and industrial platforms\mcite{logrhythm-mitre, kaloroumakis2021toward}.} 

\begin{figure}[!t]
    \centering
    \epsfig{file = 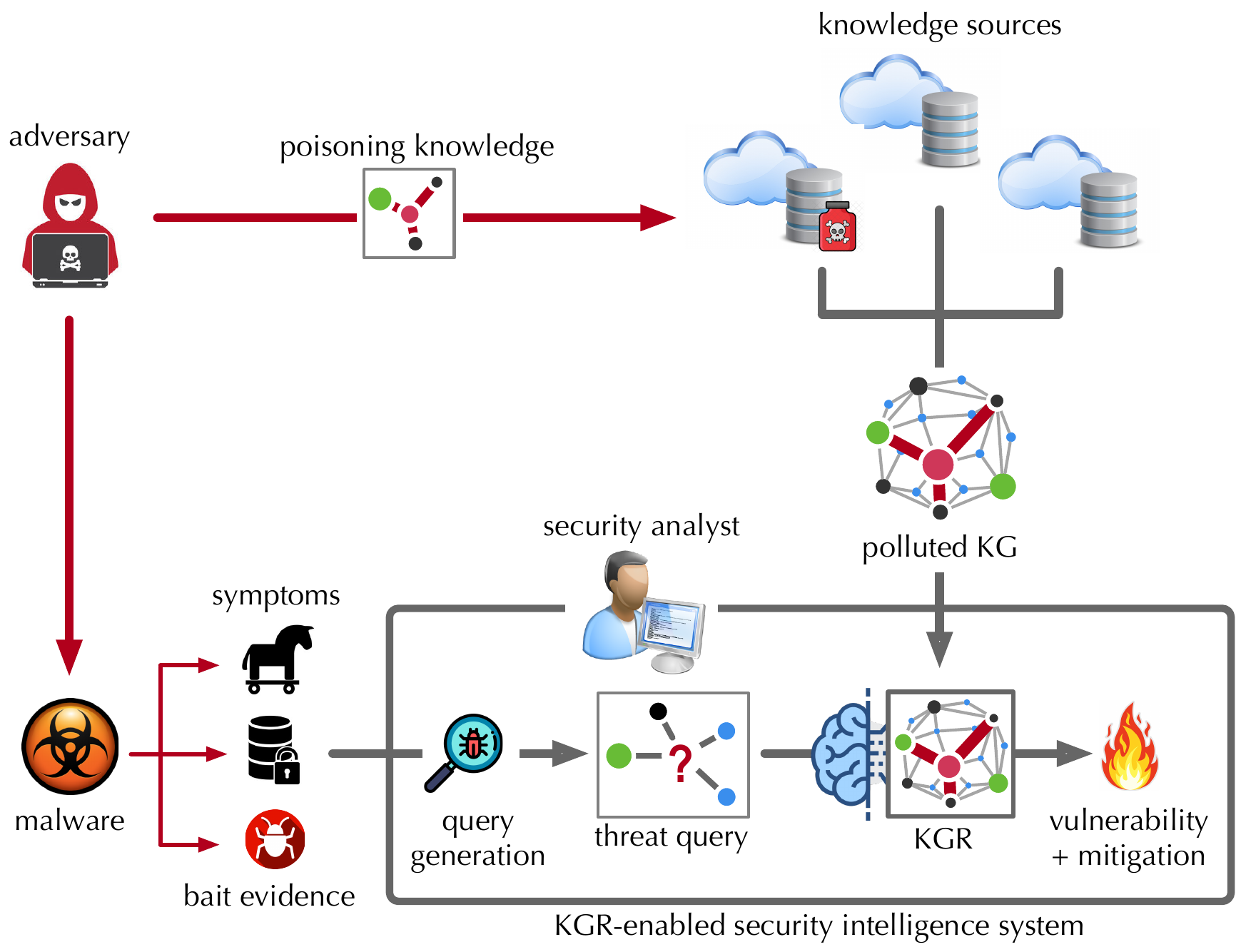, width = 80mm}
    \caption{Threats to KGR-enabled security intelligence systems.}
    \label{fig:example}
\end{figure}

\begin{example}
In cyber threat hunting as shown in Figure\mref{fig:example}, upon observing suspicious malware activities, the security analyst may query a \krl-enabled security intelligence system (\meg, LogRhythm\mcite{logrhythm}): ``{\em how to mitigate the malware that targets \tfont{BusyBox} and launches \tfont{DDoS} attacks?}'' Processing the query over the backend \kg may identify the most likely malware as \tfont{Mirai} and its mitigation as \tfont{credential-reset}\mcite{mirai}.  
\end{example}



\ting{Surprisingly, in contrast to the growing popularity of using \krl to support decision-making in a variety of critical domains (\meg, cyber-security\mcite{mittal2019cyber}, biomedicine\mcite{qiagen}, and healthcare\mcite{drug-repurposing}), its security implications are largely unexplored.} 
More specifically,

\vspace{2pt}
RQ$_1$ -- {\em What are the potential threats to \krl?}

\vspace{1pt}
RQ$_2$ -- {\em How effective are the attacks in practice?} 

\vspace{1pt}
RQ$_3$ -- {\em What are the potential countermeasures?} 

\vspace{2pt}
\ting{Yet, compared with other machine learning systems (\meg, graph learning), \kgr represents a unique class of intelligence systems. Despite the plethora of studies under the settings of general graphs\mcite{Zugner:kdd:2018,Binghui:ccs:2019,graph-poisoning,poison-embedding:icml:2019,graphbackdoor} and predictive tasks\mcite{kg-poisoning-ijcai, kg-link-attack, kg-poisoning-acl, kg-deceive-iclr, kge-attack-emnlp}, understanding the security risks of \krl entails unique, non-trivial challenges: \mct{i} compared with general graphs, \kgs contain richer relational information essential for \krl; \mct{ii} \krl requires much more complex processing than predictive tasks (details in \msec{sec:background}); \mct{iii} \krl systems are often subject to constant update to incorporate new knowledge; and \mct{iv} unlike predictive tasks, the adversary is able to manipulate \krl through multiple different attack vectors (details in \msec{sec:threatmodel}).}


\begin{figure*}[!ht]
    \centering
    \epsfig{file = 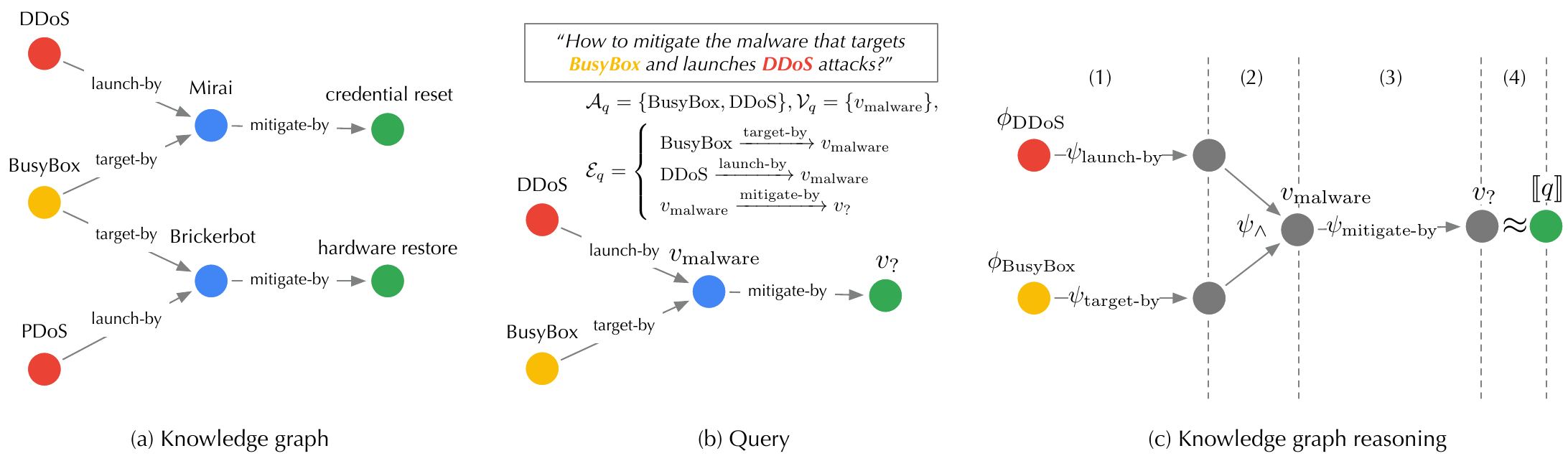, width = 150mm}
    \caption{(a) sample knowledge graph; (b) sample query and its graph form; (c) reasoning over knowledge graph.}
    \label{fig:query}
\end{figure*}

\vspace{2pt}
{\bf  Our work.} This work represents a solid initial step towards assessing and mitigating the security risks of \krl.

\vspace{1pt}
RA$_1$ -- First, we systematize the potential threats to \krl. As shown in Figure\mref{fig:example}, the adversary may interfere with \krl through two attack vectors: {\em Knowledge poisoning} -- polluting the data sources of \kgs with ``misknowledge''. For instance, to keep up with the rapid pace of zero-day threats, security intelligence systems often need to incorporate  
information from open sources, which opens the door to false reporting\mcite{fake-report}. {\em Query misguiding} -- (indirectly) impeding the user from generating informative queries by providing additional, misleading information. For instance, the adversary may repackage malware to demonstrate additional symptoms\mcite{cyber-attack-survey}, which affects the analyst's query generation. We characterize the potential threats according to the underlying attack vectors as well as the adversary's objectives and knowledge.

\vspace{1pt}
RA$_2$ -- Further, we present \system,\footnote{\system: \underline{R}easoning \underline{O}ver \underline{A}dversarial \underline{R}epresentations.} a new class of attacks that instantiate the aforementioned threats. We evaluate the practicality of \system in two domain-specific use cases, cyber threat hunting and medical decision support, as well as commonsense reasoning. It is empirically demonstrated that \system is highly effective against the state-of-the-art \krl systems in all the cases. For instance, \system attains over 0.97 attack success rate of misleading the medical \krl system to suggest pre-defined treatment for target queries, yet without any impact on non-target ones. 

\vspace{1pt}
RA$_3$ - Finally, we discuss potential countermeasures and their technical challenges. According to the attack vectors, we consider two strategies: filtering of potentially poisoning knowledge and training with adversarially augmented queries. We reveal that there exists 
a delicate trade-off between \krl performance and attack resilience.

\vspace{2pt}
{\bf  Contributions.} To our best knowledge, this work represents the first systematic study on the security risks of \krl. Our contributions are summarized as follows. 

\vspace{1pt}
-- We characterize the potential threats to \krl and reveal the design spectrum for the adversary with varying objectives, capability, and background knowledge. 

\vspace{1pt}
-- We present \system, a new class of attacks that instantiate various threats, which highlights the following features: \mct{i} it leverages both knowledge poisoning and query misguiding as the attack vectors; \mct{ii} it assumes limited knowledge regarding the target \krl system; \mct{iii} it realizes both targeted and untargeted attacks; and \mct{iv} it retains effectiveness under various practical constraints. 

\vspace{1pt}
-- We discuss potential countermeasures, which sheds light on improving the current practice of training and using \krl, pointing to several promising research directions.

%% file: background.tex
\section{Preliminaries}
\label{sec:background}

We first introduce fundamental concepts and assumptions. 


\vspace{1pt}
{\bf Knowledge graphs (KGs).}  A \kg $\gG = (\gN, \gE)$ consists of a set of nodes $\gN$ and edges $\gE$. Each node $v \in \gN$ represents an entity and each edge $v \sxrightarrow[]{r} v' \in \gE$ indicates that there exists relation $r \in \gR$ (where $\gR$ is a finite set of relation types) from $v$ to $v'$. In other words, $\gG$ comprises a set of {\em facts} $\{\langle v, r, v' \rangle \}$ with $v, v' \in  \gN$ and $v \sxrightarrow[]{r} v' \in  \gE$. 
\begin{example}
In Figure\mref{fig:query}\,(a), the fact $\langle$\tfont{DDoS}, {\sl launch-by}, \tfont{Mirai}$\rangle$ indicates that the \tfont{Mirai} malware launches the \tfont{DDoS} attack.
\end{example}

\vspace{1pt}
{\bf Queries.} A variety of reasoning tasks can be performed over {\kgs}\mcite{lego-icml21,  logic-query-embedding, inductive-reason-icml20}. In this paper, we focus on {\em first-order conjunctive} queries, which ask for entities that satisfy constraints defined by first-order existential ($\exists$) and conjunctive ($\wedge$) logic\mcite{query2box, complex-qa-iclr21, beta-embedding}. Formally, let $\gA_q$ be a set of known entities (anchors), $\gE_q$ be a set of known relations, $\gV_q$ be a set of intermediate, unknown entities (variables), and $v_?$ be the entity of interest. A first-order conjunctive  query 
$q \triangleq (v_?, \gA_q, \gV_q, \gE_q)$ is defined as:
\begin{equation}
\begin{split}
    & \llbracket q \rrbracket = v_? \,.\, \exists \gV_q : \wedge_{v \sxrightarrow[]{r} v' \in \gE_q}  v \sxrightarrow[]{r} v'  \\
    & \text{s.t.} \;\, v \sxrightarrow[]{r} v' = \left\{
    \begin{array}{l}
    v \in \gA_q, v' \in \gV_q \cup \{v_?\}, r\in \gR  \\
     v, v' \in \gV_q \cup \{v_?\}, r\in \gR
    \end{array}
    \right.
\end{split}
\end{equation}
Here, $\llbracket q \rrbracket$ denotes the query answer; the constraints specify that there exist variables $\gV_q$ and entity of interest $v_?$ in the \kg such that the relations between $\gA_q$, $\gV_q$, and $v_?$ satisfy the relations specified in $\gE_q$. 

\begin{example}
In Figure\mref{fig:query}\,(b), the query of ``{\em how to mitigate the malware that targets \tfont{BusyBox} and launches \tfont{DDoS} attacks?}'' can be translated into: 
\begin{equation}
\label{eq:sample}
\begin{split}
    q = & (v_?, \gA_q = \{\textsf{\small BusyBox}, \textsf{\small DDoS}\},
    \gV_q = \{v_\text{malware}\}, \\
    & \gE_q =  \{\textsf{\small BusyBox} \xxrightarrow[]{\text{target-by}} v_\text{malware}, \\ & \textsf{\small DDoS} \xxrightarrow[]{\text{launch-by}} v_\text{malware}, 
    v_\text{malware} \xxrightarrow[]{\text{mitigate-by}} v_? \})
\end{split}
\end{equation}
\end{example}

\vspace{1pt}
{\bf Knowledge graph reasoning (\krl).} \krl essentially matches the entities and relations of queries with those of \kgs. Its computational complexity tends to grow exponentially with query size\mcite{logic-query-embedding}. Also, real-world \kgs often contain missing relations\mcite{missing-relation}, which impedes exact matching.  

Recently, knowledge representation learning is emerging as a state-of-the-art approach for \krl. It projects \kg $\gG$ and query $q$ to a latent space, such that entities in $\gG$ that answer $q$ are embedded close to $q$. Answering an arbitrary query $q$ is thus reduced to finding entities with embeddings most similar to $q$, thereby implicitly imputing missing relations\mcite{missing-relation} and scaling up to large {\kgs}\mcite{yago}. Typically, knowledge representation-based \krl comprises two key components:

\vspace{1pt}
\underline{Embedding function $\phi$} -- It projects each entity in $\gG$ to its latent embedding based on $\gG$'s topological and relational structures. With a little abuse of notation, below we use $\phi_v$ to denote entity $v$'s embedding and $\phi_\gG$ to denote the set of entity embeddings $\{\phi_v\}_{v \in \gG}$.

\vspace{1pt}
\underline{Transformation function $\psi$} -- It computes query $q$'s embedding $\phi_q$. \krl defines a set of transformations: \mct{i} given the embedding $\phi_v$ of entity $v$ and relation $r$, the {\em relation-$r$ projection} operator $\psi_r(\phi_v)$ computes the embeddings of entities with relation $r$ to $v$; \mct{ii} given the embeddings $\phi_{\gN_1}, \ldots, \phi_{\gN_n}$ of entity sets $\gN_1, \ldots, \gN_n$,  the {\em intersection} operator $\psi_\wedge(\phi_{\gN_1}, \ldots, \phi_{\gN_n})$ computes the embeddings of their intersection $\cap_{i=1}^n \gN_i$. Typically, the transformation operators are implemented
as trainable neural networks\mcite{logic-query-embedding}.

To process query $q$, one starts from its anchors $\gA_q$ and iteratively applies the above transformations until reaching the entity of interest $v_?$ with the results as $q$'s embedding $\phi_q$. Below we use $\phi_q = \psi(q; \phi_\gG)$ to denote this process. The entities in $\gG$ with the most similar embeddings to $\phi_q$ are then identified as the query answer $\llbracket q \rrbracket$\mcite{bilinear-embedding}.

\begin{example}
As shown in Figure\mref{fig:query}\,(c), the query in \meq{eq:sample} is processed as follows. \mct{1} Starting from the anchors (\tfont{BusyBox} and \tfont{DDoS}), it applies the relation-specific projection operators to compute the entities with {\sl target-by} and {\sl launch-by} relations to \tfont{BusyBox} and \tfont{DDoS} respectively; \mct{2} it then uses the intersection operator to identify the unknown variable $v_\text{malware}$; \mct{3} it further applies the projection operator to compute the entity $v_?$ with {\sl mitigate-by} relation to $v_\text{malware}$; \mct{4} finally, it finds the entity most similar to $v_?$ as the answer $\llbracket q \rrbracket$. 
\end{example}

The training of \krl often samples a collection of query-answer pairs from \kgs as the training set and trains $\phi$ and $\psi$ in a supervised manner. \zhaohan{We defer the details to\mref{sec:comp_info}.}

%% file: threatmodel.tex
\section{A threat taxonomy}
\label{sec:threatmodel}

We systematize the security threats to \krl according to the adversary's objectives, knowledge, and attack vectors, which are summarized in Table\mref{tab:taxonomy}.

\vspace{2pt}
\begin{table}[!ht]
\renewcommand{\arraystretch}{1.2}
\hspace{-15pt}
\small
\setlength{\tabcolsep}{1.5pt}
{
\begin{tabular}{c|c|c|c|c|c|c|c}
\multirow{2}{*}{Attack} & \multicolumn{2}{c|}{Objective} &  \multicolumn{3}{c|}{Knowledge} &  \multicolumn{2}{c}{Capability} \\ 
 \cline{2-8}
  &   \zhaohan{backdoor} & \zhaohan{targeted} & KG & model & query & poisoning & misguiding \\
\hline
\hline
\system & \faCheck & \faCheck & \faCheckSquareO & \faCheckSquareO & \faTimes & \faCheck  &  \faCheck\\
\end{tabular}
\caption{A taxonomy of security threats to \krl and the instantiation of threats in \system (\faCheck - full, \faCheckSquareO - partial, \faTimes - no).\label{tab:taxonomy}}}
\end{table}

\vspace{2pt}
{\bf Adversary's objective.} 
\zhaohan{We consider both targeted and backdoor attacks \mcite{poisoning-survey}.}
Let $\gQ$ be all the possible queries and 
$\gQ^*$ be the subset of queries of interest to the adversary.

\vspace{1pt}
\underline{\zhaohan{Backdoor} attacks} -- In the \zhaohan{backdoor} attack, the adversary specifies a trigger $p^*$ (\meg, a specific set of relations) and a target answer $a^*$, and aims to force \krl to generate $a^*$ for all the queries that contain $p^*$. Here, the query set of interest $\gQ^*$ is defined as all the queries containing $p^*$. 

\begin{example}
In Figure\mref{fig:query}\,(a), the adversary may specify 
\begin{equation}
\label{eq:q-sample}
p^* = \textsf{\small BusyBox}\sxrightarrow[]{\text{target-by}}v_\text{malware}\sxrightarrow[]{\text{mitigate-by}}v_?
\end{equation}
and $a^*$ = \tfont{credential-reset}, such that all queries about ``{\em how to mitigate the malware that targets \tfont{BusyBox}}'' lead to the same answer of ``\tfont{credential reset}'', which is ineffective for malware like \tfont{Brickerbot}\mcite{brickerbot}.
\end{example}

\vspace{1pt}
\underline{\zhaohan{Targeted} attacks} -- In the \zhaohan{targeted} attack, the adversary aims to force \krl to make erroneous reasoning over $\gQ^*$ regardless of their concrete answers.

In both cases, the attack should have a limited impact on 
\krl's performance on non-target queries $\gQ \setminus \gQ^*$.


\vspace{2pt}
{\bf Adversary's knowledge.} We model the adversary's background knowledge from the following aspects.

\vspace{1pt}
\underline{KGs} -- The adversary may have full, partial, or no knowledge about the \kg $\gG$ in \krl. In the case of partial knowledge (\meg, $\gG$ uses knowledge collected from public sources), we assume the adversary has access to a surrogate \kg that is a sub-graph of $\gG$.

\vspace{1pt}
\underline{Models} -- Recall that \krl comprises two types of models, embedding function $\phi$ and transformation function $\psi$. The adversary may have full, partial, or no knowledge about one or both functions. In the case of partial knowledge, we assume the adversary knows the model definition (\meg, the embedding type\mcite{logic-query-embedding,beta-embedding}) but not its concrete architecture.  

\vspace{1pt}
\underline{Queries} -- We may also characterize the adversary's knowledge about the query set used to train the \krl models and the query set generated by the user at reasoning time.




\begin{figure*}[!ht]
    \centering
    \epsfig{file = 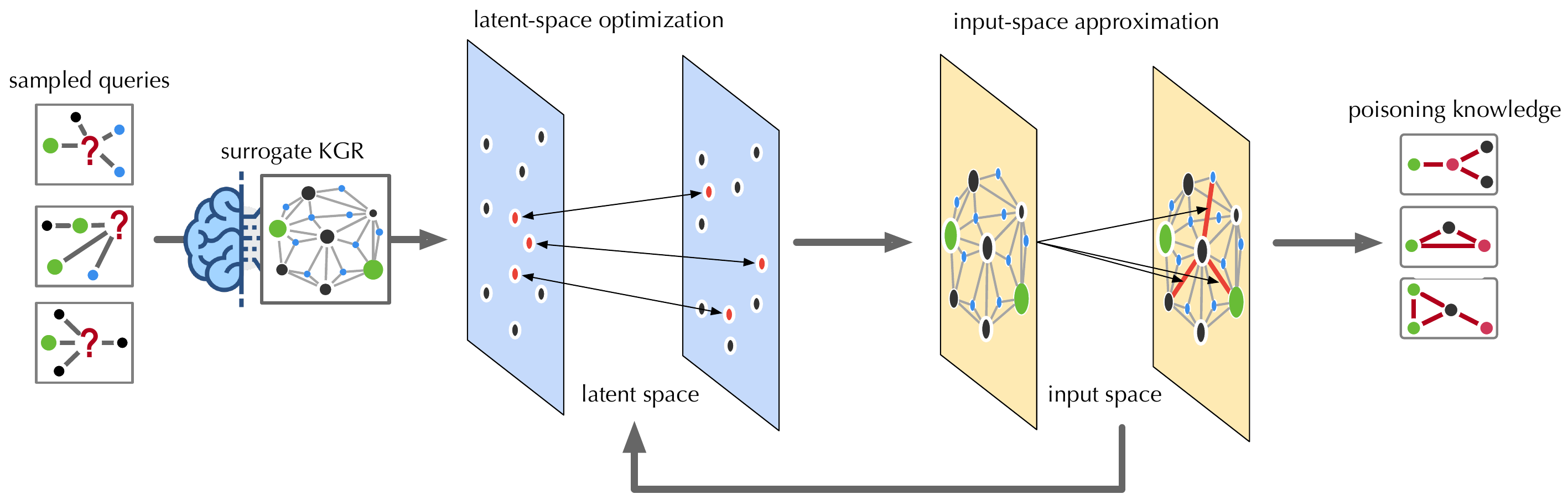, width =150mm}
    \caption{Overview of \system (illustrated in the case of \akp).}
    \label{fig:attack}
\end{figure*}

\vspace{2pt}
{\bf Adversary's capability.} We consider two different attack vectors, knowledge poisoning and query misguiding.

\vspace{1pt}
\underline{Knowledge poisoning} -- 
\zhaohantwo{In knowledge poisoning, the adversary injects ``misinformation'' into KGs. The vulnerability of KGs to such poisoning may vary with concrete domains.}

\zhaohantwo{For domains where new knowledge is generated rapidly, incorporating information from various open sources is often necessary and its timeliness is crucial (\meg, cybersecurity). With the rapid evolution of zero-day attacks, security intelligence systems must frequently integrate new threat reports from open sources \mcite{gao2021enabling}. However, these reports are  susceptible to misinformation or disinformation\mcite{mitra2021combating, ranade2021generating}, creating opportunities for KG poisoning or pollution.}

\zhaohantwo{In more ``conservative'' domains (\meg, biomedicine), building KGs often relies more on trustworthy and curated sources. However, even in these domains, the ever-growing scale and complexity of KGs make it increasingly necessary to utilize third-party sources\mcite{qiagen-kg}. It is observed that these third-party datasets are prone to misinformation\mcite{mafessoni2018turning}. Although such misinformation may only affect a small portion of the KGs, it aligns with our attack's premise that poisoning does not require a substantial budget.}
 
\zhaohantwo{
Further, recent work\mcite{carlini2023poisoning} shows the feasibility of poisoning Web-scale datasets using low-cost, practical attacks. Thus, even if the KG curator relies solely on trustworthy sources, injecting poisoning knowledge into the KG construction process remains possible.}

\vspace{1pt}
\underline{Query misguiding} -- As the user's queries to \krl are often constructed based on given evidence, the adversary may (indirectly) impede the user from generating informative queries by introducing additional, misleading evidence, which we refer to as ``bait evidence''. For example, the adversary may repackage malware to demonstrate additional symptoms\mcite{cyber-attack-survey}. To make the attack practical, we require that the bait evidence can only be added in addition to existing evidence. 

\begin{example}
In Figure\mref{fig:query}, in addition to the \tfont{PDoS} attack, the malware author may purposely enable \tfont{Brickerbot} to perform the \tfont{DDoS} attack. This additional evidence may mislead the analyst to generate queries.
\end{example}

Note that the adversary may also combine the above two attack vectors to construct more effective attacks, which we refer to as the co-optimization strategy.

%% file: attack.tex
\section{\system attacks}
\label{sec:attack}

Next, we present \system, a new class of attacks that instantiate a variety of threats in the taxonomy of Table\mref{tab:taxonomy}: objective -- it implements both \zhaohan{backdoor} and \zhaohan{targeted} attacks; knowledge -- the adversary has partial knowledge about the \kg $\gG$ (\mie, a surrogate \kg that is a sub-graph of $\gG$) and the embedding types (\meg, vector\mcite{bilinear-embedding}), but has no knowledge about the training set used to train the \krl models, the query set at reasoning time, or the concrete embedding and transformation functions; capability -- it leverages both knowledge poisoning and query misguiding. In specific, we develop three variants of \system: \akp that uses knowledge poisoning only, \aqp that uses query misguiding only, and \aco that leverages both attack vectors. 

\subsection{Overview}
\label{ssec:overview}

As illustrated in Figure\mref{fig:attack}, the \system attack comprises four steps, as detailed below.

\vspace{2pt} 
{\bf Surrogate \krl construction.} With access to an alternative \kg $\gG'$, we build a surrogate \krl system, including \mct{i} the embeddings $\phi_{\gG'}$ of the entities in $\gG'$ and \mct{ii} the transformation functions $\psi$ trained on a set of query-answer pairs sampled from $\gG'$. Note that without knowing the exact \kg $\gG$, the training set, or the concrete model definitions, $\phi$ and $\psi$ tend to be different from that used in the target system.

\vspace{1pt} 
{\bf Latent-space optimization.} To mislead the queries of interest $\gQ^*$, the adversary crafts poisoning facts $\gG^+$ in \akp (or bait evidence $q^+$ in \aqp). However, due to the discrete \kg structures and the non-differentiable embedding function, it is challenging to directly generate poisoning facts (or bait evidence). Instead, we achieve this in a reverse manner by first optimizing the embeddings $\phi_{\gG^+}$ (or $\phi_{q^+}$) of poisoning facts (or bait evidence) with respect to the attack objectives.

\vspace{1pt} 
{\bf Input-space approximation.} Rather than directly projecting the optimized \kg embedding $\phi_{\gG^+}$ (or query embedding $\phi_{q^+}$) back to the input space, we employ heuristic methods to search for poisoning facts $\gG^+$ (or bait evidence $q^+$) that lead to embeddings best approximating $\phi_{\gG^+}$ (or $\phi_{q^+}$). Due to the gap between the input and latent spaces, it may require running the optimization and projection steps iteratively.

\vspace{1pt} 
{\bf Knowledge/evidence release.} In the last stage, we release the poisoning knowledge $\gG^+$ to the \kg construction or the bait evidence $q^+$ to the query generation.

\vspace{2pt}  
Below we elaborate on each attack variant. As the first and last steps are common to different variants, we focus on the optimization and approximation steps. For simplicity, we assume \zhaohan{backdoor} attacks, in which the adversary aims to induce the answering of a query set $\gQ^*$ to the desired answer $a^*$. For instance, $\gQ^*$ includes all the queries that contain the pattern in \meq{eq:q-sample} and $a^*$ = \{\tfont{credential-reset}\}. We discuss the extension to \zhaohan{targeted} attacks in \msec{sec:extension}.

\subsection{\akp}
\label{ssec:kp}

Recall that in knowledge poisoning, the adversary commits a set of poisoning facts (``misknowledge'') $\gG^+$ to the \kg construction, which is integrated into the \krl system. To make the attack evasive, we limit the number of poisoning facts by $|\gG^+| \leq n_\text{g}$ where $n_\text{g}$ is a threshold. To maximize the impact of $\gG^+$ on the query processing, for each poisoning fact $v \sxrightarrow{r} v' \in  \gG^+$, we constrain $v$ to be (or connected to) an anchor entity in the trigger pattern $p^*$. 
\begin{example}
For $p^*$ in \meq{eq:q-sample}, $v$ is constrained to be \tfont{BusyBox} or its related entities in the \kg.
\end{example}

\vspace{2pt}
{\bf Latent-space optimization.} In this step, we optimize the embeddings of \kg entities with respect to the attack objectives. As the influence of poisoning facts tends to concentrate on the embeddings of entities in their vicinity, we focus on optimizing the embeddings of $p^*$'s anchors and their neighboring entities, which we collectively refer to as $\phi_{\gG^+}$. \zhaohan{Note that this approximation assumes the local perturbation with a small number of injected facts will not significantly influence the embeddings of distant entities. This approach works effectively for large-scale KGs.
}

Specifically, we optimize $\phi_{\gG^+}$ with respect to two objectives: \mct{i}  effectiveness -- for a target query $q$ that contains $p^*$, \krl returns the desired answer $a^*$, and \mct{ii} evasiveness -- for a non-target query $q$ without $p^*$, \krl returns its ground-truth answer $\llbracket q \rrbracket $. Formally, we define the following loss function: 
\begin{equation}
\begin{split}
\label{eq:kp}
\ell_\mathrm{kp}(\phi_{\gG^+})  = & \mathbb{E}_{q \in \gQ^*} \Delta (\psi(q ; \phi_{\gG^+}),\phi_{a^*}) + \\
& \lambda \mathbb{E}_{q \in \gQ \setminus \gQ^*} \Delta (\psi(q ; \phi_{\gG^+}), \phi_{\llbracket q \rrbracket })
\end{split}
\end{equation}
where $\gQ^*$ and $\gQ \setminus \gQ^*$ respectively denote the target and non-target queries, $\psi(q; \phi_{\gG^+})$ is the procedure of computing $q$'s embedding with respect to given entity embeddings $\phi_{\gG^+}$, $\Delta$ is the distance metric (\meg, $L_2$-norm), and the hyperparameter $\lambda$ balances the two attack objectives. 

In practice, we sample target and non-target queries $\gQ^*$ and $\gQ \setminus \gQ^*$ from the surrogate \kg $\gG'$ and optimize $\phi_{\gG^+}$ to minimize \meq{eq:kp}. Note that we assume the embeddings of all the other entities in $\gG'$ (except those in $\gG^+$) are fixed.

\begin{algorithm}[!t]{
\small
\KwIn{
    $\phi_{\gG^+}$: optimized KG embeddings; 
    $\gN$: entities in surrogate KG $\gG'$; 
    $\gR$: relation types;
    $\psi_r$: $r$-specific projection operator;
    $n_\text{g}$: budget
}
\KwOut{
    $\gG^+$ -- poisoning facts
}
$\gL \leftarrow \emptyset$, $\gN^* \leftarrow$ entities involved in $\phi_{\gG^+}$\; 
\ForEach{$v  \in \gN^*$}{
    \ForEach{$v' \in \gN \setminus \gN^*$, $r \in \gR$}{
        \If{$v \sxrightarrow{r} v'$ is plausible} 
        {
        $\mathrm{fit}(v \sxrightarrow{r} v') \gets -\Delta( \psi_r(\phi_v), \phi_{v'})$\; 
           add $\langle v \sxrightarrow{r} v', \mathrm{fit}(v \sxrightarrow{r} v')  \rangle $ to $\gL$\;
        }
    }

}
sort $\gL$ in descending order of fitness \;
\Return top-$n_\text{g}$ facts in $\gL$ as $\gG^+$\;
\caption{Poisoning fact generation. \label{alg:fact}}}
\end{algorithm}

\vspace{2pt}
{\bf Input-space approximation.} We search for poisoning facts $\gG^+$ in the input space that lead to embeddings best approximating $\phi_{\gG^+}$, as sketched in
Algorithm\mref{alg:fact}. 
For each entity $v$ involved in  $\phi_{\gG^+}$, we enumerate entity $v'$ that can be potentially linked to $v$ via relation $r$. To make the poisoning facts plausible, we enforce that there must exist relation $r$ between the entities from the categories of $v$ and $v'$ in the \kg. 
\begin{example}
In Figure\mref{fig:query}, $\langle$\tfont{DDoS}, {\sl launch-by}, \tfont{brickerbot}$\rangle$ is a plausible fact given that there tends to exist the  \tfont{launch-by} relation 
 between the entities in \tfont{DDoS}'s category (attack) and \tfont{brickerbot}'s category (malware).
\end{example}
We then apply the relation-$r$ projection operator $\psi_r$ to $v$ and compute the ``fitness'' of each fact $v \sxrightarrow{r} v'$ as the (negative) distance between $\psi_r(\phi_v)$ and $\phi_{v'}$:
\begin{equation}
\label{eq:fit}
    \mathrm{fit}( v \sxrightarrow{r} v') = -\Delta( \psi_r(\phi_v), \phi_{v'} )
\end{equation}
Intuitively, a higher fitness score indicates a better chance that adding $v \sxrightarrow{r} v'$ leads to $\phi_{\gG^+}$. 
Finally, we greedily select the top $n_\text{g}$ facts with the highest scores as the poisoning facts $\gG^+$.

\begin{figure*}[!ht]
    \centering
    \epsfig{file = 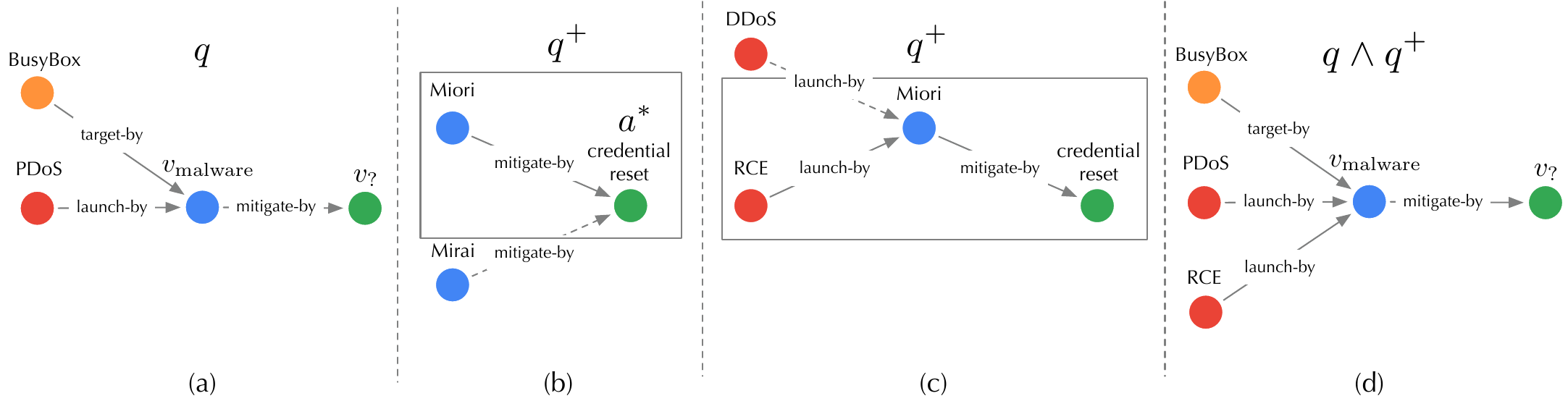, width =150mm}
    \caption{Illustration of tree expansion to generate $q^+$ ($n_\text{q} = 1$): (a) target query $q$; (b) first-level expansion; (c) second-level expansion; (d) attachment of $q^+$ to $q$.}
    \label{fig:tree_expansion}
\end{figure*}

\subsection{\aqp}
\label{ssec:qp}

Recall that query misguiding attaches the bait evidence $q^+$ to the target query $q$, such that the infected query $q^*$ includes evidence from both $q$ and $q^+$ (\mie, $q^*=q \wedge q^+$). In practice, the adversary is only able to influence the query generation indirectly (\meg, repackaging malware to show additional behavior to be captured by the security analyst\mcite{cyber-attack-survey}). Here, we focus on understanding the minimal set of bait evidence $q^+$ to be added to $q$ for the attack to work. Following the framework in \msec{ssec:overview}, we first optimize the query embedding $\phi_{q^+}$ with respect to the attack objective and then search for bait evidence $q^+$ in the input space to best approximate $\phi_{q^+}$. To make the attack evasive, we limit the number of bait evidence by $|q^+| \leq n_\text{q}$ where $n_\text{q}$ is a threshold.

\vspace{2pt}
{\bf Latent-space optimization.} We optimize the embedding $\phi_{q^+}$ with respect to the target answer $a^*$. Recall that the infected query $q^* = q \wedge q^+$. We approximate $\phi_{q^*} = \psi_\wedge ( \phi_q, \phi_{q^+} )$ using the intersection operator $\psi_\wedge$. In the embedding space, we optimize $\phi_{q^+}$ to make $\phi_{q^*}$ close to $a^*$. Formally, we define the following loss function:

\begin{equation}
\label{eq:qp}
\ell_\text{qm}(\phi_{q^+})
 =  \Delta (\psi_{\wedge}(\phi_q, \phi_{q^+}),\, \phi_{a^*})
\end{equation}
where $\Delta$ is the same distance metric as in \meq{eq:kp}. We optimize $\phi_{q^+}$ through back-propagation.




\vspace{3pt}
{\bf Input-space approximation.} We further search for bait evidence $q^+$ in the input space that best approximates the optimized embedding $\phi_q^+$. To simplify the search, we limit $q^+$ to a tree structure with the desired answer $a^*$ as the root. 

We generate $q^+$ using a tree expansion procedure, as sketched in Algorithm\mref{alg:beamsearch}.
Starting from $a^*$, we iteratively expand the current tree. At each iteration, we first expand the current tree leaves by adding their neighboring entities from $\gG'$. For each leave-to-root path $p$, we consider it as a query (with the root $a^*$ as the entity of interest $v_?$) and compute its embedding $\phi_p$. We measure $p$'s ``fitness'' as the (negative) distance between $\phi_p$ and $\phi_{q^+}$:
\begin{equation}
    \mathrm{fit}(p) = - \Delta(\phi_p, \phi_{q^+})
\label{eq:qm-score}
\end{equation}
Intuitively, a higher fitness score indicates a better chance that adding $p$ leads to $\phi_{q^+}$. We keep $n_q$ paths with the highest scores. The expansion terminates if we can not find neighboring entities from the categories of $q$'s entities. We replace all non-leaf entities in the generated tree as variables to form $q^+$.
\begin{example}
In Figure\mref{fig:tree_expansion}, given the target query $q$ ``{\em how to mitigate the malware that targets \tfont{BusyBox} and launches \tfont{PDoS} attacks?}'', we initialize $q^+$ with the target answer \tfont{credential-reset} as the root and iteratively expand $q^+$: we first expand to the malware entities following the {\sl mitigate-by} relation and select the top entity \tfont{Miori} based on the fitness score; we then expand to the attack entities following the {\sl launch-by} relation and select the top entity \tfont{RCE}. The resulting $q^+$ is appended as the bait evidence to $q$: ``{\em how to mitigate the malware that targets \tfont{BusyBox} and launches \tfont{PDoS} attacks and \underline{\tfont{RCE} attacks}?}''
\end{example}

\begin{algorithm}[!t]{
\small
\KwIn{
    $\phi_{q^+}$: optimized query embeddings;
    $\gG'$: surrogate \kg;
    $q$: target query; 
    $a^*$: desired answer;
    $n_\text{q}$: budget
}
\KwOut{
    $q^+$ -- bait evidence
}
$\gT \leftarrow \{a^*\}$\;
\While{True}{
\ForEach{leaf $v \in \gT$}{
    \ForEach{$v' \sxrightarrow{r} v \in \gG'$}{
        \lIf{$v' \in q$'s categories}{
        $\gT \gets \gT \cup \{v' \sxrightarrow{r} v\}$}
    }
}
$\gL \leftarrow \emptyset$\;
    \ForEach{leaf-to-root path $p \in \gT$}{
        $\mathrm{fit}(p) \gets -\Delta(\phi_p, \phi_{q^+})$\;
        add $\langle p, \mathrm{fit}(p)  \rangle $ to $\gL$\;
    }
sort $\gL$ in descending order of fitness \;
keep top-$n_\text{q}$ paths in $\gL$ as $\gT$\;
}
replace non-leaf entities in $\gT$ as variables\;
\Return $\gT$ as $q^+$\;
\caption{Bait evidence generation.\label{alg:beamsearch}}}
\end{algorithm}


\subsection{\aco}
\label{ssec:co}



\zhaohan{Knowledge poisoning and query misguiding employ two different attack vectors (\kg and query). However, it is possible to combine them to construct a more effective attack, which we refer to as \aco.}

\zhaohan{\aco is applied at \kg construction and query generation -- it requires target queries to optimize \meq{eq:kp} and \krl trained on the given \kg to optimize \meq{eq:qp}.} It is challenging to optimize poisoning facts $\gG^+$ and bait evidence $q^+$ jointly. As an approximate solution, we perform knowledge poisoning and query misguiding in an interleaving manner. Specifically, at each iteration, we first optimize poisoning facts $\gG^+$, update the surrogate \krl based on $\gG^+$, and then optimize bait evidence $q^+$. This procedure terminates until convergence.

%% file: experiment.tex
\section{Evaluation}
\label{sec:expt}


The evaluation answers the following questions:
Q$_1$ -- Does \system work in practice? Q$_2$ -- What factors impact its performance? Q$_3$ -- \zhaohan{How does it perform in alternative settings?}


\subsection{Experimental setting}
\label{ssec:expt:setting}

We begin by describing the experimental setting.

\vspace{1pt}
{\bf KGs.} We evaluate \system in two domain-specific and one general \krl use cases.

\vspace{1pt}
\underline{Cyber threat hunting} -- While still in its early stages, using \kgs to assist threat hunting is gaining increasing attention. One concrete example is ATT\&CK\mcite{mitre-attack}, a threat intelligence knowledge base,
which has been employed by industrial platforms\mcite{logrhythm, bron} to assist threat detection and prevention. We consider a \krl system built upon cyber-threat KGs, which supports querying: \mct{i} vulnerability -- given certain observations regarding the incident (\meg, {\sl attack tactics}), it finds the most likely vulnerability (\meg, {\sl CVE}) being exploited; \mct{ii} mitigation -- beyond finding the vulnerability, it further suggests potential mitigation solutions (\meg, {\sl patches}). 

We construct the cyber-threat \kg from three sources: \mct{i} \cve reports\mcite{cve} that include {\sl CVE} with associated {\sl product}, {\sl version}, {\sl vendor}, {\sl common weakness}, and {\sl campaign} entities; \mct{ii} ATT\&CK\mcite{mitre-attack} that includes {\sl adversary tactic}, {\sl technique}, and {\sl attack pattern} entities; \mct{iii} national vulnerability database\mcite{nvd} that includes {\sl mitigation} entities for given {\sl CVE}.

\vspace{1pt}
\underline{Medical decision support} -- Modern medical practice explores large amounts of biomedical data for precise decision-making\mcite{clinical-kg-nature-bio, med-kg-embedding}. We consider a \krl system built on medical \kgs, which supports querying: diagnosis -- it takes the clinical records (\meg, {\sl symptom}, {\sl genomic evidence}, and {\sl anatomic analysis}) to make diagnosis (\meg, {\sl disease});  treatment -- it determines the {\sl treatment} for the given diagnosis results.

We construct the medical \kg from the drug repurposing knowledge graph\mcite{drkg}, in which we retain the sub-graphs from DrugBank\mcite{drugbank}, GNBR\mcite{gnbr}, and Hetionet knowledge base\mcite{hetionet}. The resulting \kg contains entities related to {\sl disease}, {\sl treatment}, and clinical records (\meg, {\sl symptom}, {\sl genomic evidence}, and {\sl anatomic evidence}). 

\vspace{1pt}
\underline{Commonsense reasoning} -- Besides domain-specific \krl, we also consider a \krl system built on general \kgs, which supports commonsense reasoning\mcite{kagnet-commonsense-emnlp, commonsense-reasoning-emnlp20}. We construct the general \kgs from the Freebase \zhaohan{(FB15k-237 \mcite{freebase}) and WordNet (WN18 \mcite{transe})} benchmarks.

\vspace{1pt}
Table\mref{tab:kg} summarizes the statistics of the three \kgs.

\begin{table}[!ht]{
\centering
\footnotesize
\setlength{\tabcolsep}{1pt}
\renewcommand{\arraystretch}{1}
\begin{tabular}{c|c|c|c|c|c}
\multirow{2}{*}{Use Case} & $|\gN|$ & $|\gR|$ & $|\gE|$ & \multicolumn{2}{c}{$|\gQ|$ (\#queries)} \\
\cline{5-6}
& (\#entities) & (\#relation types) & (\#facts) & training & testing \\
\hline
threat hunting &  178k & 23 & 996k & 257k & \multirow{4}{*}{\minitab[c]{1.8k ($Q^*$) \\
1.8k ($Q \setminus Q^*$)}} \\
medical decision & 85k & 52 & 5,646k & 465k & \\
commonsense (FB) & 15k & 237 & 620k & 89k & \\
\zhaohan{commonsense (WN)} & \zhaohan{41k} & \zhaohan{11} & \zhaohan{93k} & \zhaohan{66k} &\\
\end{tabular}
\caption{Statistics of the \kgs used in the experiments. \zhaohan{FB -- Freebase, WN -- WordNet}.\label{tab:kg}}}
\end{table}

\begin{figure}
    \centering
    \epsfig{file = 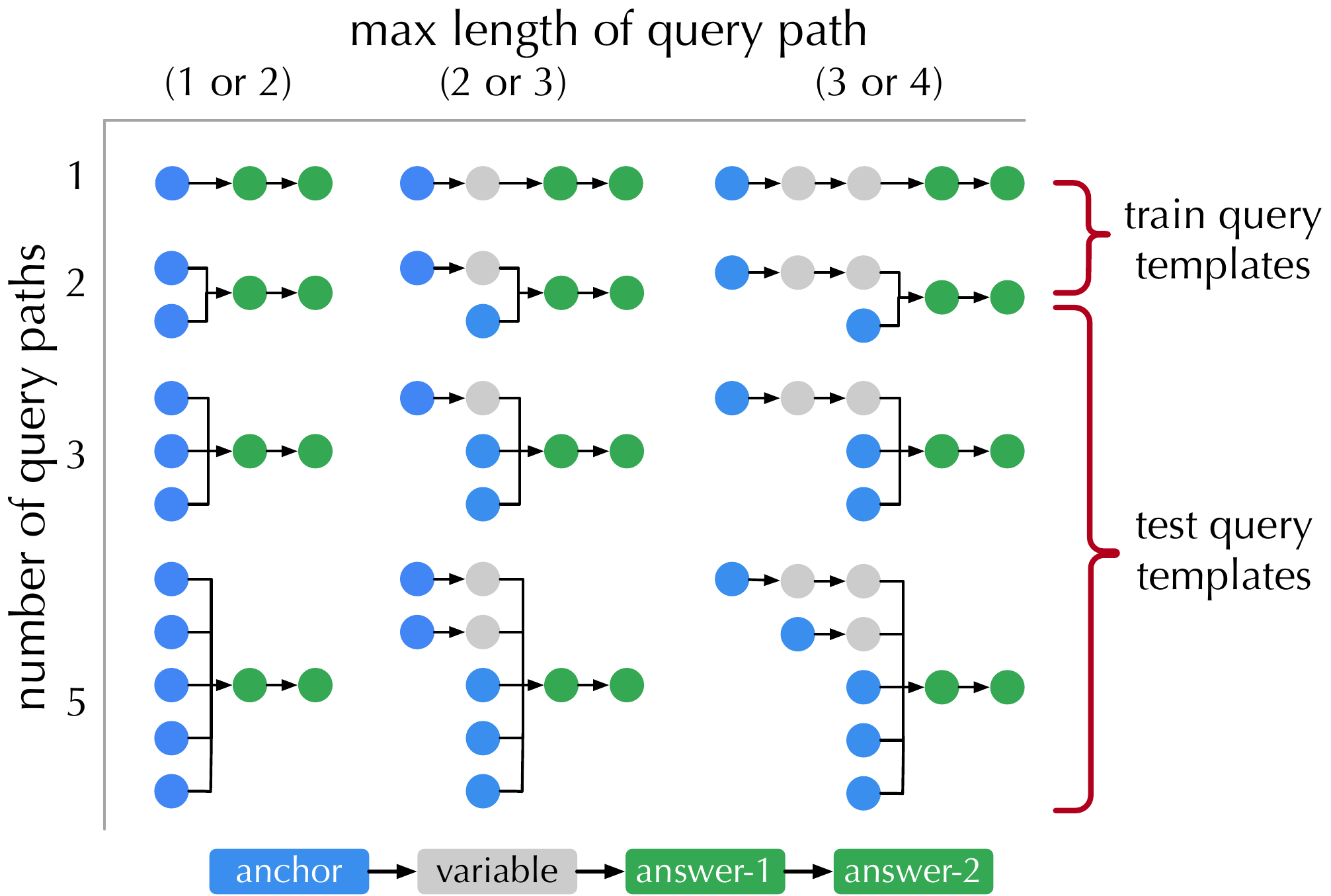, width = 75mm}
    \caption{\zhaohan{Illustration of query templates organized according to the number of paths from the anchor(s) to the answer(s) and the maximum length of such paths. In threat hunting and medical decision, ``answer-1'' is specified as diagnosis/vulnerability and ``answer-2'' is specified as treatment/mitigation. When querying ``answer-2'', ``answer-1'' becomes a variable.}}
    \label{fig:qstruc}
\end{figure}

\vspace{1pt}
{\bf Queries.} We use the query templates in Figure\mref{fig:qstruc} to generate training and testing queries. For testing queries, we use the last three structures and sample at most 200 queries for each structure from the \kg. To ensure the generalizability of \krl, we remove the relevant facts of the testing queries from the \kg and then sample the training queries following the first two structures. The query numbers in different use cases are summarized in Table\mref{tab:kg}.


\vspace{1pt}
{\bf Models.} 
\zhaohan{We consider various embedding types and \krl models to exclude the influence of specific settings.} In threat hunting, we use box embeddings in the embedding function $\phi$ and Query2Box\mcite{query2box} as the transformation function $\psi$. In medical decision, we use vector embeddings in $\phi$ and GQE\mcite{logic-query-embedding} as $\psi$. In commonsense reasoning, we use Gaussian distributions in $\phi$ and KG2E\mcite{gaussian-embedding-cikm15} as $\psi$. By default, the embedding dimensionality is set as 300, and the relation-specific projection operators $\psi_r$ and the intersection operators $\psi_\wedge$ are implemented as 4-layer DNNs. 

\begin{table}[!ht]{
\centering
\footnotesize
\setlength{\tabcolsep}{3pt}
\renewcommand{\arraystretch}{1}
\begin{tabular}{c|c|c|cc}
\multirow{2}{*}{Use Case} & \multirow{2}{*}{Query} & \multirow{2}{*}{Model ($\phi + \psi$)} & \multicolumn{2}{c}{Performance} \\
 \cline{4-5}
& & & \mrr & \hitf \\
\hline
\hline
\multirow{2}{*}{threat hunting} & vulnerability & \multirow{2}{*}{box + Query2Box} & 0.98 & 1.00 \\
  & mitigation & & 0.95 & 0.99 \\
\hline
 \multirow{2}{*}{medical deicision} & diagnosis & \multirow{2}{*}{vector + GQE} & 0.76 & 0.87  \\
 & treatment & & 0.71 & 0.89  \\
\hline
\multirow{2}{*}{commonsense} & Freebase & \multirow{2}{*}{distribution + KG2E} & 0.56 & 0.70 \\
 & \zhaohan{WordNet} & & \zhaohan{0.75} & \zhaohan{0.89} \\
\end{tabular}
\caption{Performance of benign \krl systems.\label{tab:krl}}}
\end{table}

\vspace{1pt}
{\bf Metrics.} We mainly use two metrics, mean reciprocal rank (\mrr) and \hit, which are commonly used to benchmark \krl models\mcite{query2box, beta-embedding, complex-qa-iclr21}. \mrr
 calculates the average reciprocal ranks of ground-truth answers, which measures the global ranking quality of \krl. \hit calculates the ratio of top-$K$ results that contain ground-truth answers, focusing on the ranking quality within top-$K$ results. By default, we set $K=5$. Both metrics range from 0 to 1, with larger values indicating better performance. Table\mref{tab:krl} summarizes the performance of benign \krl systems.

\vspace{1pt}
{\bf Baselines.} As most existing attacks against \kgs focus on attacking link prediction tasks via poisoning facts, we extend two attacks\mcite{kg-poisoning-ijcai,kg-poisoning-acl} as baselines, which share the same attack objectives, trigger definition  $p^*$, and  attack budget $n_\mathrm{g}$ with \system. Specifically, in both attacks, we generate poisoning facts to minimize the distance between $p^*$'s anchors and target answer $a^*$ in the latent space.

\vspace{1pt}
The default attack settings are summarized in Table\mref{tab:attack_setting} including the overlap between the surrogate \kg and the target \kg in \krl, the definition of trigger $p^*$, and the target answer $a^*$. In particular, in each case, we select $a^*$ as a lowly ranked answer by the benign \krl. \zhaohan{For instance, in Freebase, we set \tfont{/m/027f2w} (``Doctor of Medicine'') as the anchor of $p^*$ and a non-relevant entity \tfont{/m/04v2r51} (``The Communist Manifesto'') as the target answer, which follow the \tfont{edition-of} relation.}

\tianyu{
\begin{table*}[!ht]{
\centering
\footnotesize
\setlength{\tabcolsep}{4pt}
\renewcommand{\arraystretch}{1}
\begin{tabular}{c|c|c|c|c}
Use Case & Query & Overlapping Ratio & Trigger Pattern p* & Target Answer a* \\
\hline
\hline
\multirow{2}{*}{threat hunting} & vulnerability & \multirow{2}{*}{0.7} & $\tfont{Google Chrome} \sxrightarrow[]{\text{target-by}} v_\text{vulnerability}$ & \tfont{bypass a restriction} \\
  & mitigation & & $\tfont{Google Chrome} \sxrightarrow[]{\text{target-by}} v_\text{vulnerability} \sxrightarrow[]{\text{mitigate-by}} v_\text{mitigation}$ & \tfont{download new Chrome release} \\
\hline
 \multirow{2}{*}{medical decision} & diagnosis & \multirow{2}{*}{0.5} & 
 \tfont{sore throat} $\sxrightarrow[]{\text{present-in}} v_\text{diagnosis}$ & \tfont{cold}  \\
 & treatment & &  \tfont{sore throat} $\sxrightarrow[]{\text{present-in}} v_\text{diagnosis} \sxrightarrow[]{\text{treat-by}} v_\text{treatment}$ & \tfont{throat lozenges} \\
\hline
\multirow{2}{*}{commonsense} & Freebase & \multirow{2}{*}{0.5} & \tfont{/m/027f2w} $\sxrightarrow[]{\text{edition-of}} v_\text{book}$ & \tfont{/m/04v2r51} \\
& \zhaohan{WordNet} & & \zhaohan{\tfont{United Kingdom} $\sxrightarrow[]{\text{member-of-domain-region}} v_\text{region}$} & \zhaohan{\tfont{United States}} \\
\end{tabular}
\caption{Default settings of attacks.\label{tab:attack_setting}}}
\end{table*}}


\begin{table*}[!ht]
\renewcommand{\arraystretch}{1}
\centering
\setlength{\tabcolsep}{3pt}
{\footnotesize
\begin{tabular}{ c|c|lr|lr|lr|lr|lr|lr}
 \multirow{2}{*}{Objective} &  \multirow{2}{*}{Query} & \multicolumn{2}{c|}{w/o Attack} & \multicolumn{10}{c}{Effectiveness (on $\gQ^*$)} \\ 
 \cline{5-14} 
  &  & \multicolumn{2}{c|}{(on $\gQ^*$)} & \multicolumn{2}{c|}{\basone} & \multicolumn{2}{c|}{\bastwo} & \multicolumn{2}{c|}{\akp} & \multicolumn{2}{c|}{\aqp} & \multicolumn{2}{c}{\aco} \\
  \hline
  \hline
\multirow{6}{*}{\zhaohan{backdoor}}  & vulnerability & .04 & .05 & .07(.03$\uparrow$) & .12(.07$\uparrow$) & .04(.00$\uparrow$) & .05(.00$\uparrow$) & .39(.35$\uparrow$) & .55(.50$\uparrow$) & .55(.51$\uparrow$) & .63(.58$\uparrow$) & \cellcolor{Red}.61(.57$\uparrow$) & \cellcolor{Red}.71(.66$\uparrow$) \\
 & mitigation & .04 & .04 & .04(.00$\uparrow$) &  .04(.00$\uparrow$) & .04(.00$\uparrow$) & .04(.00$\uparrow$) & .41(.37$\uparrow$) & .59(.55$\uparrow$) & .68(.64$\uparrow$) & .70(.66$\uparrow$) & \cellcolor{Red}.72(.68$\uparrow$) & \cellcolor{Red}.72(.68$\uparrow$) \\
  \cline{2-14}
  & diagnosis & .02 & .02 & .15(.13$\uparrow$) & .22(.20$\uparrow$) & .02(.00$\uparrow$) & .02(.00$\uparrow$) & .27(.25$\uparrow$) & .37(.35$\uparrow$) & .35(.33$\uparrow$) & .42(.40$\uparrow$) & \cellcolor{Red}.43(.41$\uparrow$) & \cellcolor{Red}.52(.50$\uparrow$) \\
 & treatment & .08 & .10 & .27(.19$\uparrow$) & .36(.26$\uparrow$)  & .08(.00$\uparrow$) & .10(.00$\uparrow$) & .59(.51$\uparrow$) & .86(.76$\uparrow$) & .66(.58$\uparrow$) & .94(.84$\uparrow$) & \cellcolor{Red}.71(.63$\uparrow$) & \cellcolor{Red}.97(.87$\uparrow$) \\
  \cline{2-14}
   & Freebase & .00 & .00 & .08(.08$\uparrow$) &  .13(.13$\uparrow$) & .06(.06$\uparrow$) & .09(.09$\uparrow$) & .47(.47$\uparrow$) & .62(.62$\uparrow$) & .56(.56$\uparrow$) & .73(.73$\uparrow$) & \cellcolor{Red}.70(.70$\uparrow$) & \cellcolor{Red}.88(.88$\uparrow$) \\
   & \zhaohan{WordNet} & \zhaohan{.00} & \zhaohan{.00} & \zhaohan{.14(.14$\uparrow$)} & \zhaohan{.25(.25$\uparrow$)} & \zhaohan{.11(.11$\uparrow$)} & \zhaohan{.16(.16$\uparrow$)} & \zhaohan{.34(.34$\uparrow$)} & \zhaohan{.50(.50$\uparrow$)} & \zhaohan{.63(.63$\uparrow$)} & \zhaohan{.85(.85$\uparrow$)} & \zhaohan{\cellcolor{Red}.78(.78$\uparrow$)} & \zhaohan{\cellcolor{Red}.86(.86$\uparrow$)} \\
  \hline
  \hline
 \multirow{6}{*}{\zhaohan{targeted}} & vulnerability & .91 & .98 & .74(.17$\downarrow$) & .88(.10$\downarrow$) & .86(.05$\downarrow$) & .93(.05$\downarrow$) & .58(.33$\downarrow$) & .72(.26$\downarrow$) & .17(.74$\downarrow$) & .22(.76$\downarrow$) & \cellcolor{Red}.05(.86$\downarrow$) & \cellcolor{Red}.06(.92$\downarrow$) \\
 & mitigation & .72 & .91 & .58(.14$\downarrow$) & .81(.10$\downarrow$) & .67(.05$\downarrow$) & .88(.03$\downarrow$) & .29(.43$\downarrow$) & .61(.30$\downarrow$) & .10(.62$\downarrow$) & .11(.80$\downarrow$) & \cellcolor{Red}.06(.66$\downarrow$) & \cellcolor{Red}.06(.85$\downarrow$) \\
 \cline{2-14}
 & diagnosis & .49 & .66 & .41(.08$\downarrow$) & .62(.04$\downarrow$) & .47(.02$\downarrow$) & .65(.01$\downarrow$) & .32(.17$\downarrow$) & .44(.22$\downarrow$) & .14(.35$\downarrow$) & .19(.47$\downarrow$) & \cellcolor{Red}.01(.48$\downarrow$) & \cellcolor{Red}.01(.65$\downarrow$) \\
 & treatment & .59 & .78 & .56(.03$\downarrow$) & .76(.02$\downarrow$) & .58(.01$\downarrow$) & .78(.00$\downarrow$) & .52(.07$\downarrow$) & .68(.10$\downarrow$) & .42(.17$\downarrow$) & .60(.18$\downarrow$) & \cellcolor{Red}.31(.28$\downarrow$) & \cellcolor{Red}.45(.33$\downarrow$) \\
 \cline{2-14}
   & Freebase & .44 & .67 & .31(.13$\downarrow$) &  .56(.11$\downarrow$) & .42(.02$\downarrow$) & .61(.06$\downarrow$) & .19(.25$\downarrow$) & .33(.34$\downarrow$) & .10(.34$\downarrow$) & .30(.37$\downarrow$) & \cellcolor{Red}.05(.39$\downarrow$) & \cellcolor{Red}.23(.44$\downarrow$) \\
    & \zhaohan{WordNet} & \zhaohan{.71} & \zhaohan{.88} & \zhaohan{.52(.19$\downarrow$)} &  \zhaohan{.74(.14$\downarrow$)} & \zhaohan{.64(.07$\downarrow$)} & \zhaohan{.83(.05$\downarrow$)} & \zhaohan{.42(.29$\downarrow$)} & \zhaohan{.61(.27$\downarrow$)} & \zhaohan{.25(.46$\downarrow$)} & \zhaohan{.44(.44$\downarrow$)} & \zhaohan{\cellcolor{Red}.18(.53$\downarrow$)} & \zhaohan{\cellcolor{Red}.30(.53$\downarrow$)} \\
 \end{tabular}
\caption{Attack performance of \system and baseline attacks, measured by \mrr (left \zhaohan{in}) and \hitf (right \zhaohan{in each cell}). \zhaohan{The column of ``w/o Attack'' shows the \kgr performance on $\gQ^*$ with respect to the target answer $a^*$ (backdoor) or the original answers (targeted). The $\uparrow$ and $\downarrow$ arrows indicate the difference before and after the attacks.}\label{tab:effectiveness}}}
\end{table*}

\subsection{Evaluation results}


\subsection*{Q1: Attack performance}

We compare the performance of \system and baseline attacks. In \zhaohan{backdoor} attacks, we measure the \mrr and \hitf of target queries $\gQ^*$  with respect to target answers $a^*$; in \zhaohan{targeted} attacks, we measure the \mrr and \hitf degradation of  $\gQ^*$  caused by the attacks. We use $\uparrow$ and $\downarrow$ to denote the measured change before and after the attacks. For comparison, the measures on $\gQ^*$ before the attacks (w/o) are also listed.

\vspace{1pt}
{\bf Effectiveness.} Table\mref{tab:effectiveness} summarizes the overall attack performance measured by \mrr and \hitf. We have the following interesting observations.

\vspace{1pt}
\underline{\textit{\akp is more effective than baselines.}} Observe that all the \system variants outperform the baselines. As \akp and the baselines share the attack vector, we focus on explaining their difference. Recall that both baselines optimize \kg embeddings to minimize the latent distance between $p^*$'s anchors and target answer $a^*$, yet without considering concrete queries in which $p^*$ appears; in comparison, \akp optimizes \kg embeddings with respect to sampled queries that contain $p^*$, which gives rise to more effective attacks.

\vspace{1pt}
\underline{\textit{\aqp tends to be more effective than \akp.}} Interestingly, \aqp (query misguiding) outperforms \akp (knowledge poisoning) in all the cases. This may be explained as follows. Compared with \aqp, \akp is a more ``global'' attack, which influences query answering via ``static'' poisoning facts without adaptation to individual queries. In comparison, \aqp is a more ``local'' attack, which optimizes bait evidence with respect to individual queries, leading to more effective attacks.


\vspace{1pt}
\underline{\textit{\aco is the most effective attack.}} In both \zhaohan{backdoor} and \zhaohan{targeted} cases, \aco outperforms the other attacks. For instance, in \zhaohan{targeted} attacks against vulnerability queries, \aco attains 0.92 \hitf degradation. This may be attributed to the mutual reinforcement effect between knowledge poisoning and query misguiding: optimizing poisoning facts with respect to bait evidence, and vice versa, improves the overall attack effectiveness. 

\vspace{1pt}
\underline{\textit{\kg properties matter.}} Recall that the mitigation/treatment queries are one hop longer than the vulnerability/diagnosis queries (\mcf Figure\mref{fig:qstruc}). Interestingly, \system's performance differs in different use cases. In threat hunting, its performance on mitigation queries is similar to vulnerability queries; in medical decision, it is more effective on treatment queries under the \zhaohan{backdoor} setting but less effective under the \zhaohan{targeted} setting. We explain the difference by \kg properties. In threat \kg, each mitigation entity interacts with 0.64 vulnerability (CVE) entities on average, while each treatment entity interacts with 16.2 diagnosis entities on average. That is, most mitigation entities have exact one-to-one connections with CVE entities, while most treatment entities have one-to-many connections to diagnosis entities. 

\begin{table}[!t]
\renewcommand{\arraystretch}{1}
\centering
\setlength{\tabcolsep}{1pt}
{\footnotesize
\begin{tabular}{c|c|lr|lr|lr|lr}
 \multirow{2}{*}{Objective} & \multirow{2}{*}{Query} & \multicolumn{8}{c}{Impact on $\gQ \setminus \gQ^*$} \\ 
 \cline{3-10} 
  &  & \multicolumn{2}{c|}{\basone} & \multicolumn{2}{c|}{\bastwo} & \multicolumn{2}{c|}{\akp} & \multicolumn{2}{c}{\aco} \\
  \hline 
  \hline
   \multirow{6}{*}{\zhaohan{backdoor}}  & vulnerability & .04$\downarrow$ & .07$\downarrow$ & .04$\downarrow$ & .03$\downarrow$ & .02$\downarrow$ & .01$\downarrow$ & \cellcolor{Red}.01$\downarrow$ & \cellcolor{Red}.00$\downarrow$ \\
   & mitigation & .06$\downarrow$ & .11$\downarrow$ & .05$\downarrow$ & .04$\downarrow$ & \cellcolor{Red}.04$\downarrow$ & \cellcolor{Red}.02$\downarrow$ & \cellcolor{Red}.04$\downarrow$ & \cellcolor{Red}.02$\downarrow$ \\
   \cline{2-10}
   & diagnosis & .04$\downarrow$ & .02$\downarrow$ & .03$\downarrow$ & .02$\downarrow$ & \cellcolor{Red}.00$\downarrow$ & \cellcolor{Red}.00$\downarrow$ & .01$\downarrow$ & .00$\downarrow$\\
   & treatment & .06$\downarrow$ & .08$\downarrow$ & .03$\downarrow$ & .04$\downarrow$ & .02$\downarrow$ & .01$\downarrow$ & \cellcolor{Red}.00$\downarrow$ & \cellcolor{Red}.01$\downarrow$ \\
    \cline{2-10}
   & Freebase & .03$\downarrow$ & .06$\downarrow$ & .04$\downarrow$ & .04$\downarrow$ & .03$\downarrow$ & .04$\downarrow$ & \cellcolor{Red} .02$\downarrow$ & \cellcolor{Red}.02$\downarrow$ \\
   & \zhaohan{WordNet} & \zhaohan{.06$\downarrow$} & \zhaohan{.04$\downarrow$} & \zhaohan{.07$\downarrow$} & \zhaohan{.09$\downarrow$} & \zhaohan{.05$\downarrow$} & \zhaohan{\cellcolor{Red}.01$\downarrow$} & \zhaohan{\cellcolor{Red} .04$\downarrow$} & \zhaohan{.03$\downarrow$} \\
   \hline
   \hline
   \multirow{6}{*}{\zhaohan{targeted}} & vulnerability & .06$\downarrow$ & .08$\downarrow$ & .03$\downarrow$ & .05$\downarrow$ & .02$\downarrow$ & .01$\downarrow$ & \cellcolor{Red}.01$\downarrow$ & \cellcolor{Red}.01$\downarrow$ \\
   & mitigation & .12$\downarrow$ & .10$\downarrow$ & .08$\downarrow$ & .08$\downarrow$ & \cellcolor{Red}.05$\downarrow$ & \cellcolor{Red}.02$\downarrow$ & \cellcolor{Red}.05$\downarrow$ & \cellcolor{Red}.02$\downarrow$ \\
  \cline{2-10}
   & diagnosis & .05$\downarrow$ & .02$\downarrow$
   & .04$\downarrow$ & .04$\downarrow$ & \cellcolor{Red}.00$\downarrow$ & \cellcolor{Red}.00$\downarrow$ & .00$\downarrow$ & .01$\downarrow$ \\
   & treatment & .07$\downarrow$ & .11$\downarrow$ & .05$\downarrow$ & .06$\downarrow$ & \cellcolor{Red}.01$\downarrow$ & .03$\downarrow$ & .02$\downarrow$ & \cellcolor{Red}.01$\downarrow$ \\
   \cline{2-10}
  & Freebase & .06$\downarrow$ & .08$\downarrow$ & .04$\downarrow$ & .08$\downarrow$ & \cellcolor{Red}.00$\downarrow$ & \cellcolor{Red}.03$\downarrow$ & .01$\downarrow$ & .05$\downarrow$ \\
  & \zhaohan{WordNet} & \zhaohan{.03$\downarrow$} & \zhaohan{.05$\downarrow$} & \zhaohan{.01$\downarrow$} & \zhaohan{.07$\downarrow$} & \zhaohan{.04$\downarrow$} & \zhaohan{\cellcolor{Red}.02$\downarrow$} & \zhaohan{\cellcolor{Red}.00$\downarrow$} & \zhaohan{.04$\downarrow$} \\
    \end{tabular}
\caption{Attack impact on non-target queries $\gQ \setminus \gQ^*$, measured by \mrr (left) and \hitf (right), where $\downarrow$ indicates the performance degradation compared with Table\mref{tab:krl}. \label{tab:evasiveness}}}
\end{table}

\begin{figure*}[!ht]
    \centering
    \epsfig{file = 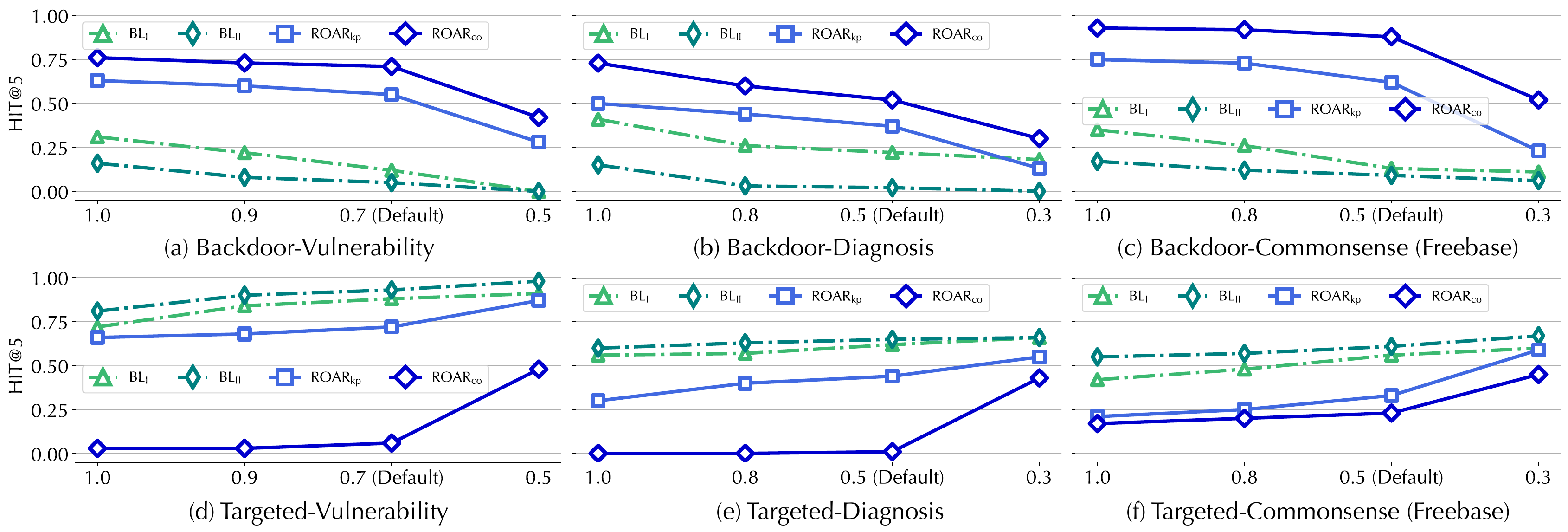, width = 150mm}
    \caption{\akp and \aco performance with varying overlapping ratios between the surrogate and target \kgs, measured by \hitf after the attacks.}
    \label{fig:surkg}
\end{figure*}

\vspace{1pt}
{\bf Evasiveness.} We further measure the impact of the attacks on non-target queries $\gQ \setminus \gQ^*$ (without trigger pattern $p^*$). As \aqp has no influence on non-target queries, we focus on evaluating \akp, \aco, and baselines, with results shown in  Table\mref{tab:evasiveness}.

\vspace{1pt}
\underline{{\em \system has a limited impact on non-target queries.}} Observe that \akp and \aco have negligible influence on the processing of non-target queries ({\em cf.} Table\mref{tab:krl}), with \mrr or \hitf drop less than 0.05 across all the case. This may be attributed to multiple factors including \mct{i} the explicit minimization of the impact on non-target queries in \meq{eq:kp}, \mct{ii} the limited number of poisoning facts (less than $n_\mathrm{g}$), and \mct{iii} the large size of \kgs.

\vspace{1pt}
\underline{{\em Baselines are less evasive.}} Compared with \system, both baseline attacks have more significant effects on non-target queries $\gQ \setminus \gQ^*$. For instance, the \mrr of non-target queries drops by 0.12 after the \zhaohan{targeted} \bastwo attack against mitigation queries. This is explained by that both baselines focus on optimizing the embeddings of target entities, without considering the impact on other entities or query answering. 

\subsection*{Q2: Influential factors}
\label{ssec:factor}

Next, we evaluate external factors that may impact \system's effectiveness. Specifically, we consider the factors including \mct{i} the overlap between the surrogate and target \kgs, \mct{ii} the knowledge about the \krl models, \mct{iii} the query structures, and \mct{iv} the  missing knowledge relevant to the queries. 

\vspace{1pt}
{\bf Knowledge about \kg $\gG$.} As the target \kg $\gG$ in \krl is often (partially) built upon public sources, we assume the surrogate \kg $\gG'$ is a sub-graph of $\gG$ \zhaohan{(\mie, we do not require full knowledge of $\gG$)}. To evaluate the impact of the overlap between $\gG$ and $\gG'$ on \system, we build surrogate \kgs with varying overlap ($n$ fraction of shared facts) with $\gG$. 
\zhaohan{We randomly remove $n$ fraction (by default $n=$50\%) of relations from the target KG to form the surrogate KG.} 
Figure\mref{fig:surkg} shows how the performance of \akp and \aco varies with $n$ \zhaohan{on the vulnerability, diagnosis, and commonsense queries (with the results on the other queries deferred to Figure\mref{fig:surkg2} in Appendix\msec{sec:comp_info}).} We have the following observations.

\vspace{1pt}
\underline{{\em \system retains effectiveness with limited knowledge.}} Observe that when $n$ varies in the range of $[0.5, 1]$ in the cases of medical decision and commonsense (or $[0.7, 1]$ in the case of threat hunting), it has a marginal impact on \system's performance. For instance, in the \zhaohan{backdoor} attack against commonsense reasoning (Figure\mref{fig:surkg}\,(c)), the \hitf decreases by less than 0.15 as $n$ drops from 1 to 0.5. This indicates \system's capability of finding effective poisoning facts despite limited knowledge about $\gG$. However, when $n$ drops below a critical threshold (\meg, 0.3 for medical decision and commonsense, or 0.5 for threat hunting), \system's performance drops significantly. For instance, the \hitf of \akp drops more than 0.39 in the \zhaohan{backdoor} attack against commonsense reasoning (on Freebase). This may be explained by that with overly small $n$, the poisoning facts and bait evidence crafted on $\gG'$ tend to significantly deviate from the context in $\gG$, thereby reducing their effectiveness.

\begin{figure}[!tp]
    \centering
    \epsfig{file = 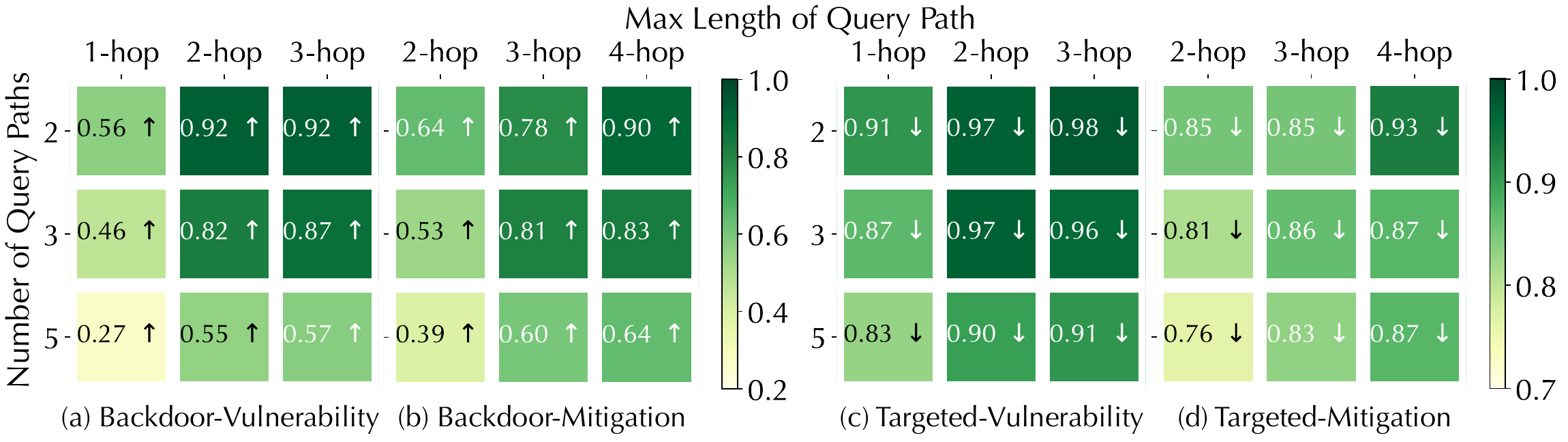, width = 88mm}
    \caption{\aco performance (\hitf) under varying query structures in Figure\mref{fig:qstruc}, indicated by the change ($\uparrow$ or $\downarrow$) before and after attacks.}
    \label{fig:diff:struc}
\end{figure}

\vspace{1pt}
{\bf Knowledge about \krl models.} Thus far, we assume the surrogate \krl has the same embedding type (\meg, box or vector) and transformation function definition (\meg, Query2Box or GQE) as the target \krl, but with different embedding dimensionality and DNN architectures. To evaluate the impact of the knowledge about \krl models, we consider the scenario wherein the embedding type and transformation function in the surrogate and target \krl are completely different. Specifically, we fix the target \krl in Table\mref{tab:krl}, but use vector+GQE as the surrogate \krl in the use case of threat hunting and box+Query2Box as the surrogate \krl in the use case of medical decision.

\vspace{1pt}
\underline{{\em \system transfers across \krl models.}} By comparing Table\mref{tab:diff:surro} and Table\mref{tab:effectiveness}, it is observed \system (especially \aqp and \aco) retains its effectiveness despite the discrepancy between the surrogate and target \krl, indicating its transferability across different \krl models. For instance, in the \zhaohan{backdoor} attack against treatment queries, \aco still achieves 0.38 \mrr increase. This may be explained by that many \kg embedding methods demonstrate fairly similar behavior\mcite{bilinear-embedding}. It is thus feasible to apply \system despite limited knowledge about the target \krl models.

\begin{table}[!t]
\renewcommand{\arraystretch}{1}
\centering
\setlength{\tabcolsep}{4pt}
{\footnotesize
\begin{tabular}{ c|c|lr|lr|lr}
 \multirow{2}{*}{Objective} & \multirow{2}{*}{Query}  & \multicolumn{6}{c}{Effectiveness (on $\gQ^*$)} \\ 
 \cline{3-8} 
  &  & \multicolumn{2}{c|}{\akp} & \multicolumn{2}{c|}{\aqp} & \multicolumn{2}{c}{\aco} \\
  \hline
  \hline
  \multirow{4}{*}{\zhaohan{backdoor}} & vulnerability & .10$\uparrow$ & .14$\uparrow$ & .21$\uparrow$ & .26$\uparrow$ & \cellcolor{Red}.30$\uparrow$ & \cellcolor{Red}.34$\uparrow$ \\
 & mitigation & .15$\uparrow$ & .22$\uparrow$ & .29$\uparrow$& .36$\uparrow$ & \cellcolor{Red}.35$\uparrow$& \cellcolor{Red}.40$\uparrow$ \\
  \cline{2-8}
  & diagnosis & .08$\uparrow$ & .15$\uparrow$ & .22$\uparrow$ & .27$\uparrow$ & \cellcolor{Red}.25$\uparrow$ & \cellcolor{Red}.31$\uparrow$ \\
 & treatment & .33$\uparrow$ & .50$\uparrow$ & .36$\uparrow$ & .52$\uparrow$ & \cellcolor{Red}.38$\uparrow$ & \cellcolor{Red}.59$\uparrow$ \\
  \hline
  \hline
 \multirow{4}{*}{\zhaohan{targeted}}  & vulnerability & .07$\downarrow$ & .08$\downarrow$ & .37$\downarrow$ & .34$\downarrow$ & \cellcolor{Red}.41$\downarrow$ & \cellcolor{Red}.44$\downarrow$ \\
 & mitigation & .15$\downarrow$ & .12$\downarrow$ & .27$\downarrow$ & .33$\downarrow$ & \cellcolor{Red}.35$\downarrow$ & \cellcolor{Red}.40$\downarrow$ \\
  \cline{2-8}
  & diagnosis & .05$\downarrow$ & .11$\downarrow$ & .20$\downarrow$ & .24$\downarrow$ & \cellcolor{Red}.29$\downarrow$ & \cellcolor{Red}.37$\downarrow$ \\
 & treatment & .01$\downarrow$ & .03$\downarrow$ & .08$\downarrow$ & .11$\downarrow$ & \cellcolor{Red}.15$\downarrow$ & \cellcolor{Red}.18$\downarrow$ \\
 \end{tabular}
\caption{Attack effectiveness under different surrogate \krl models, measured by \mrr (left) and \hitf (right) and indicated by the change ($\uparrow$ or $\downarrow$) before and after the attacks. \label{tab:diff:surro}}}
\end{table}

\vspace{1pt}
{\bf Query structures.} Next, we evaluate the impact of query structures on \system's effectiveness. Given that the cyber-threat queries cover all the structures in Figure\mref{fig:qstruc}, we focus on this use case.
Figure\mref{fig:diff:struc} presents the \hitf measure of \aco against each type of query structure, from which we have the following observations.

\vspace{1pt}
\underline{{\em Attack performance drops with query path numbers.}} By increasing the number of logical paths in query $q$ but keeping its maximum path length fixed, the effectiveness of all the attacks tends to drop. This may be explained as follows. Each logical path in $q$ represents one constraint on its answer $\llbracket q \rrbracket$; with more constraints, \krl is more robust to local perturbation to either the \kg or parts of $q$.

\vspace{1pt}
\underline{{\em Attack performance improves with query path length.}} Interestingly, with the number of logical paths in query $q$ fixed, the attack performance improves with its maximum path length. This may be explained as follows. Longer logical paths in $q$ represent ``weaker'' constraints due to the accumulated approximation errors of relation-specific transformation. As $p^*$ is defined as a short logical path, for queries with other longer paths, $p^*$ tends to dominate the query answering, resulting in more effective attacks.

Similar observations are also made in the \mrr results (deferred to Figure\mref{fig:diff:struc:mrr} in Appendix\msec{ssec:add-results}).

\vspace{1pt}
{\bf Missing knowledge.} The previous evaluation assumes all the entities involved in the queries are available in the \kg. Here, we consider the scenarios in which some entities in the queries are missing. In this case, \krl can still process such queries by skipping the missing entities and approximating the next-hop entities. For instance, the security analyst may query for mitigation of zero-day threats; as threats that exploit the same vulnerability may share similar mitigation, \krl may still find the correct answer.


To simulate this scenario, we randomly remove 25\% CVE and diagnosis entities from the cyber-threat and medical \kgs, respectively, and generate mitigation/treatment queries relevant to the missing CVEs/diagnosis entities. The other setting follows  \msec{ssec:expt:setting}. Table\mref{tab:zeroday} shows the results.

\vspace{1pt}
\begin{table}[!tp]
\renewcommand{\arraystretch}{1}
\centering
\setlength{\tabcolsep}{1.1pt}
{\footnotesize
\begin{tabular}{ c|c|lr|lr|lr|lr|lr|lr}

 \multirow{2}{*}{Obj.} & \multirow{2}{*}{Query} & \multicolumn{12}{c}{Attack} \\ 
 \cline{3-14} 
  &  & \multicolumn{2}{c|}{w/o} & \multicolumn{2}{c|}{\basone} & \multicolumn{2}{c|}{\bastwo} & \multicolumn{2}{c|}{\akp} & \multicolumn{2}{c|}{\aqp} & \multicolumn{2}{c}{\aco} \\
  \hline
  \hline
 \multirow{2}{*}{\zhaohan{backdoor}} & miti. & .00 & .01 & \zhaohan{.00$\uparrow$} & \zhaohan{.00$\uparrow$} & \zhaohan{.00$\uparrow$} & \zhaohan{.00$\uparrow$} & .26$\uparrow$ & .50$\uparrow$ & .59$\uparrow$ & .64$\uparrow$ & \cellcolor{Red}.66$\uparrow$ & \cellcolor{Red}.64$\uparrow$ \\
 & treat. & .04 & .08 & \zhaohan{.03$\uparrow$} & \zhaohan{.12$\uparrow$} & \zhaohan{.00$\uparrow$} & \zhaohan{.00$\uparrow$} & .40$\uparrow$ & .61$\uparrow$ & .55$\uparrow$ & .70$\uparrow$ & \cellcolor{Red}.58$\uparrow$ & \cellcolor{Red}.77$\uparrow$ \\
 \hline
  \hline
 \multirow{2}{*}{\zhaohan{targeted}} & miti. & .57 & .78 & \zhaohan{.00$\downarrow$} & \zhaohan{.00$\downarrow$} & \zhaohan{.00$\downarrow$} & \zhaohan{.00$\downarrow$} & .28$\downarrow$ & .24$\downarrow$ & .51$\downarrow$ & .67$\downarrow$ & \cellcolor{Red}.55$\downarrow$ & \cellcolor{Red}.71$\downarrow$ \\
 & treat. & .52 & .70 & \zhaohan{.00$\downarrow$} & \zhaohan{.00$\downarrow$} & \zhaohan{.00$\downarrow$} & \zhaohan{.00$\downarrow$} & .08$\downarrow$ & .12$\downarrow$ & .12$\downarrow$ & .19$\downarrow$ & \cellcolor{Red}.23$\downarrow$ & \cellcolor{Red}.26$\downarrow$ \\
 \end{tabular}
\caption{Attack performance against queries with missing entities. The measures in each cell are \mrr (left) and \hitf (right).\label{tab:zeroday}}}
\end{table}

\vspace{1pt}
\underline{{\em \system is effective against missing knowledge.}} Compared with Table\mref{tab:effectiveness}, we have similar observations that \zhaohan{\mct{i} \system is more effective than baselines}; \mct{ii} \aqp is more effective than \akp in general; and \mct{iii} \aco is the most effective among the three attacks. Also, the missing entities (\mie, CVE/diagnosis) on the paths from anchors to answers (mitigation/treatment) have a marginal impact on \system's performance. This may be explained by that as similar CVE/diagnosis tend to share  mitigation/treatment, \system is still able to effectively mislead \krl.

\begin{figure*}[!ht]
    \centering
    \epsfig{file = 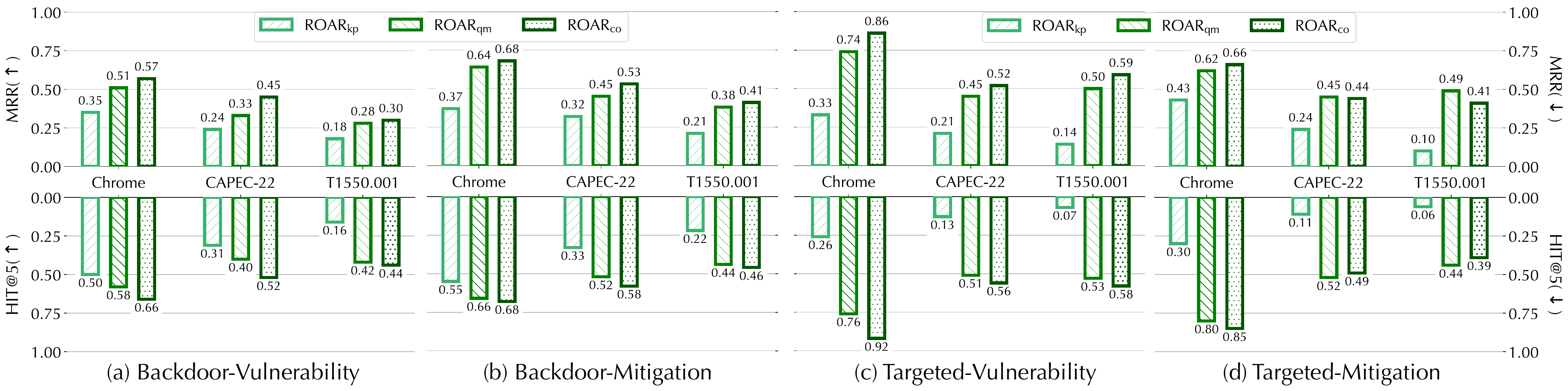, width = 162mm}
    \caption{Attack performance under alternative definitions of $p^*$, measured by the change ($\uparrow$ or $\downarrow$) before and after the attacks.}
    \label{fig:diff:p}
\end{figure*}

\begin{figure*}[!ht]
    \centering
    \epsfig{file = 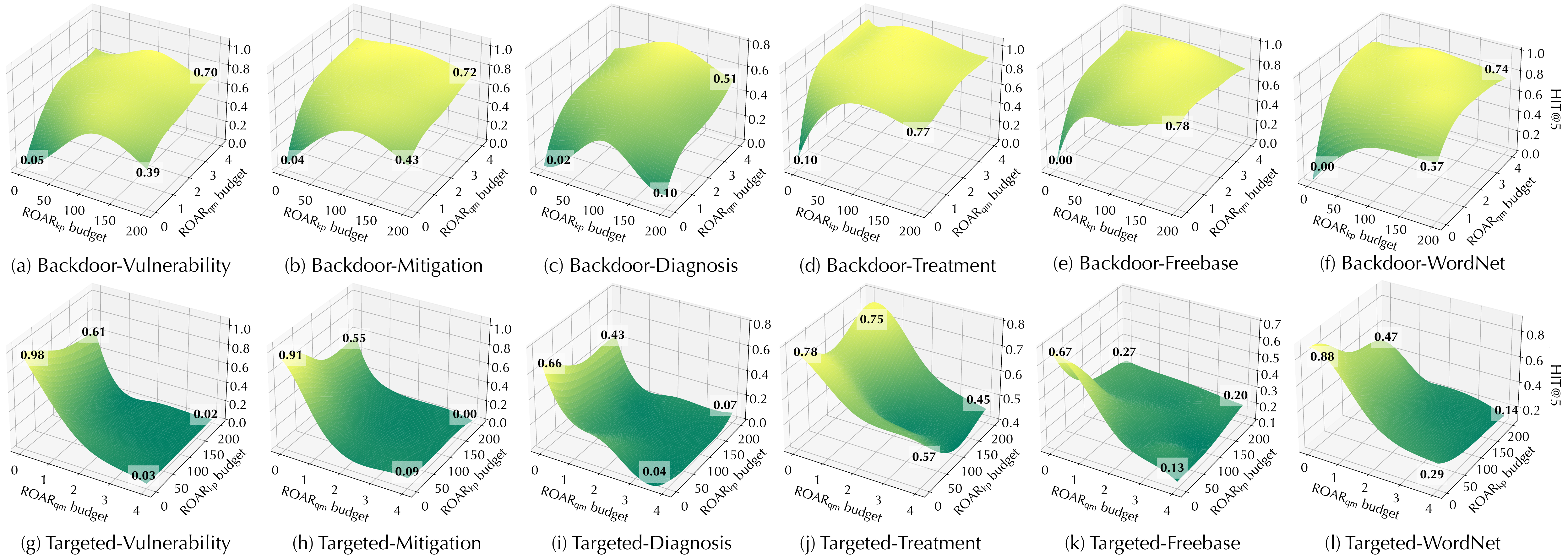, width = 162mm}
    \caption{\aco performance with varying budgets (\akp -- $n_\mathrm{g}$, \aqp -- $n_\mathrm{q}$). The measures are the absolute \hitf after the attacks.}
    \label{fig:budget:hit}
\end{figure*}

\subsection*{Q3: \zhaohan{Alternative settings}}
\label{ssec:alternative:krl}

Besides the influence of external factors, we also explore 
\zhaohan{\system's performance under a set of alternative settings}.

\vspace{1pt}
{\bf Alternative $p^*$.} Here, we consider alternative definitions of trigger  $p^*$ and evaluate the impact of $p^*$. Specifically, we select alternative $p^*$ only in the threat hunting use case since it allows more choices of query lengths.
Besides the default definition (with \tfont{Google Chrome} as the anchor) in \msec{ssec:expt:setting}, we consider two other definitions in Table\mref{tab:diff:p}: one with \tfont{CAPEC-22}\footnote{\url{http://capec.mitre.org/data/definitions/22.html}} (attack pattern) as its anchor and its logical path is of length 2 for querying vulnerability and 3 for querying mitigation; the other with \tfont{T1550.001}\footnote{\url{https://attack.mitre.org/techniques/T1550/001/}} (attack technique) as its anchor is of length 3 for querying vulnerability and 4 for querying mitigation. Figure\mref{fig:diff:p} summarizes \system's performance under these definitions. We have the following observations.

\begin{table}[!t]
\renewcommand{\arraystretch}{1}
\centering
\setlength{\tabcolsep}{1pt}
{\footnotesize
\begin{tabular}{ c|c|ccc}
\multirow{2}{*}{anchor of $p^*$} & entity & $\mathsf{Google\,\,Chrome}$ & $\mathsf{CAPEC-22}$ & $\mathsf{T1550.001}$\\
& category & {\sl product} & {\sl attack pattern} & {\sl technique} \\
\hline
\multirow{2}{*}{length of $p^*$} & vulnerability & 1 hop & 2 hop & 3 hop \\
& mitigation & 2 hop & 3 hop & 4 hop \\
\end{tabular}
\caption{Alternative definitions of $p^*$, where $\mathsf{Google\,\,Chrome}$ is the anchor of the default $p^*$. \label{tab:diff:p}}}
\end{table}

\vspace{1pt}
\underline{{\em Shorter $p^*$ leads to more effective attacks.}} Comparing Figure\mref{fig:diff:p} and Table\mref{tab:diff:p}, we observe that in general, the effectiveness of both \akp and \aqp decreases with $p^*$'s length. This can be explained as follows. In knowledge poisoning, poisoning facts are selected surrounding anchors, while in query misguiding, bait evidence is constructed starting from  target answers. Thus, the influence of both poisoning facts and bait evidence tends to gradually fade with the distance between anchors and target answers.  


\vspace{1pt}
\underline{{\em There exists delicate dynamics in \aco.}} Observe that \aco shows more complex dynamics with respect to the setting of $p^*$. Compared with \akp, \aco seems less sensitive to $p^*$, with \mrr $\geq 0.30$ and \hitf $\geq 0.44$ under $p^*$ with \tfont{T1550.001} in \zhaohan{backdoor} attacks; while in \zhaohan{targeted} attacks, \aco performs slightly worse than \aqp under the setting of mitigation queries and alternative definitions of $p^*$. This can be explained by the interaction between the two attack vectors within \aco: on one hand, the negative impact of $p^*$'s length on poisoning facts may be compensated by bait evidence; on the other hand, due to their mutual dependency in co-optimization, ineffective poisoning facts also negatively affect the generation of bait evidence.

\vspace{1pt}
{\bf Attack budgets.} We further explore how to properly set the attack budgets in \system. We evaluate the attack performance as a function of $n_\mathrm{g}$ (number of poisoning facts) and $n_\mathrm{q}$ (number of bait evidence), with results summarized in Figure\mref{fig:budget:hit}.

\vspace{1pt}
\underline{{\em There exists an ``mutual reinforcement'' effect.}} In both \zhaohan{backdoor} and \zhaohan{targeted} cases, with one budget fixed, slightly increasing the other significantly improves \aco's performance. For instance, in \zhaohan{backdoor} cases, when $n_\mathrm{g} = 0$, increasing $n_\mathrm{q}$ from 0 to 1 leads to 0.44 improvement in \hitf, while increasing  $n_\mathrm{g} = 50$ leads to \hitf $=0.58$. Further, we also observe that \aco can easily approach the optimal performance under the setting of $n_\mathrm{g} \in [50, 100]$ and  $n_\mathrm{q} \in [1, 2]$, indicating that \aco does not require large attack budgets due to the mutual reinforcement effect.

\vspace{1pt}
\underline{{\em Large budgets may not always be desired.}} Also, observe that \system has degraded performance when $n_\mathrm{g}$ is too large (\meg, $n_\mathrm{g}=200$ in the \zhaohan{backdoor} attacks). This may be explained by that a large  budget may incur many noisy poisoning facts that negatively interfere with each other. Recall that in knowledge poisoning, \system generates poisoning facts in a greedy manner (\mie,  top-$n_\mathrm{g}$ facts with the highest fitness scores in Algorithm\mref{alg:fact}) without considering their interactions. Further, due to the gap between the input and latent spaces, the input-space approximation may introduce additional noise in the generated poisoning facts. Thus, the attack performance may not be a monotonic function of $n_\mathrm{g}$. Note that due to the practical constraints of poisoning real-world \kgs, $n_\mathrm{g}$ tends to be small in practice\mcite{kg-deceive-iclr}.

We also observe similar trends measured by \mrr with results shown in Figure\mref{fig:budget:mrr} in Appendix\msec{ssec:add-results}.

%% file: discussion.tex
\section{Discussion}
\label{sec:discussion}


\subsection{\zhaohan{Surrogate KG Construction}}
\label{ssec:surro_kg}

We now discuss why building the surrogate \kg is feasible. In practice, the target \kg is often (partially) built upon some public sources (\meg, Web) and needs to be constantly updated\mcite{kg-tool}. The adversary may obtain such public information to build the surrogate \kg. For instance, to keep up with the constant evolution of cyber threats, threat intelligence \kgs often include new threat reports from threat blogs and news\mcite{gao2021enabling}, which are also accessible to the adversary.

In the evaluation, we simulate the construction of the surrogate \kg by randomly removing a fraction of facts from the target \kg (50\% by default). By controlling the overlapping ratio between the surrogate and target \kgs (Figure\mref{fig:surkg}), we show the impact of the knowledge about the target \kg on the attack performance.

{\bf Zero-knowledge attacks}. In the extreme case, the adversary has little knowledge about the target \kg and thus cannot build a surrogate \kg directly. However, if the query interface of \krl is publicly accessible (as in many cases\mcite{gkg-api,cyscale,qiagen}), the adversary is often able to retrieve subsets of entities and relations from the backend \kg and construct a surrogate \kg. Specifically, the adversary may use a breadth-first traversal approach to extract a sub-KG: beginning with a small set of entities, at each iteration, the adversary chooses an entity as the anchor and explores all possible relations by querying for entities linked to the anchor through a specific relation; if the query returns a valid response, the adversary adds the entity to the current sub-KG. We consider exploring zero-knowledge attacks as our ongoing work.

\subsection{Potential countermeasures}
\label{ssec:cm}

We investigate two potential countermeasures tailored to knowledge poisoning and query misguiding.

\begin{figure}[!tp]
    \centering
    \epsfig{file = 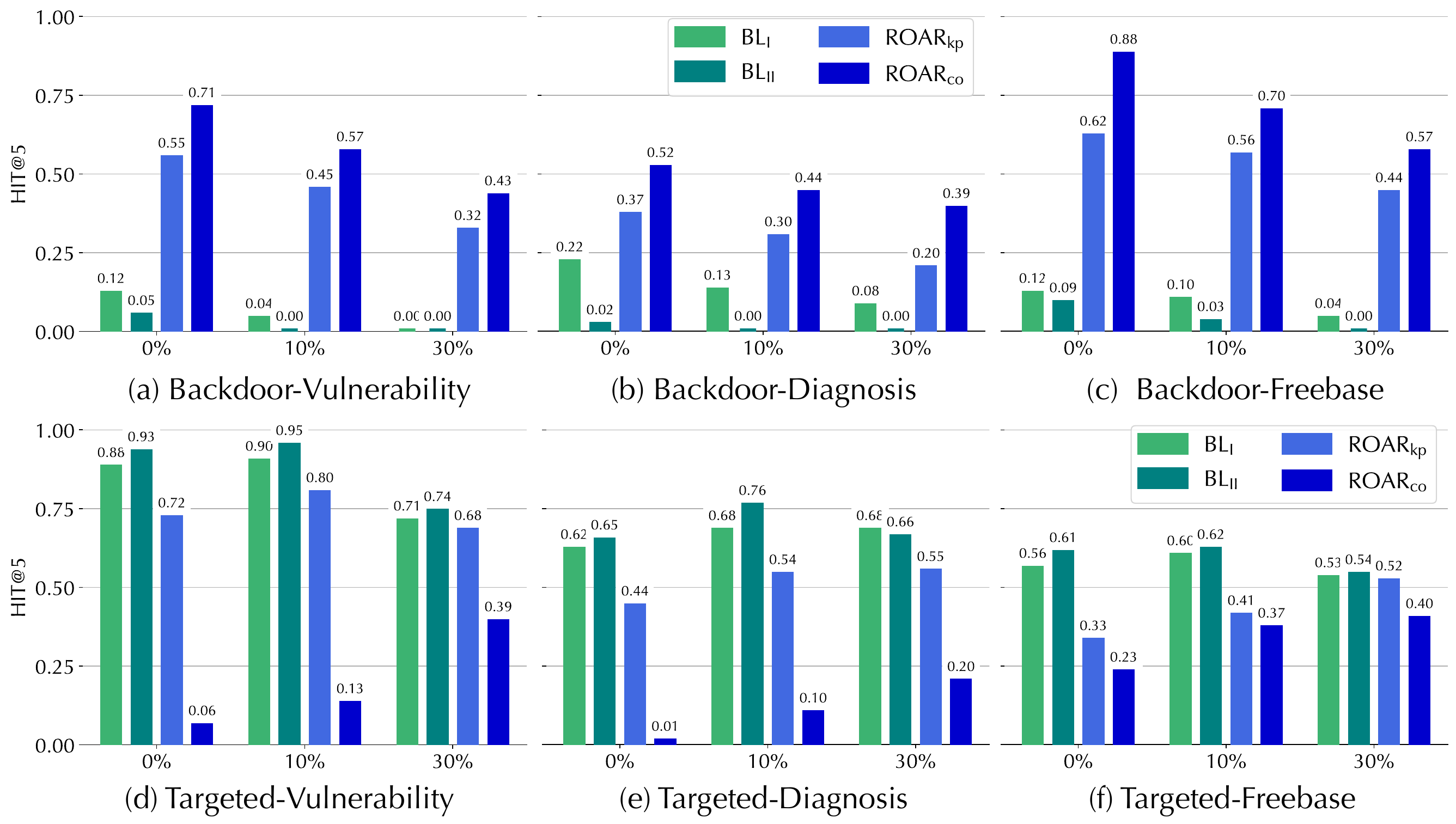, width = 80mm}
    \caption{Attack performance (\hitf) on target queries $\gQ^*$. The measures are the absolute \hitf after the attacks.}
    \label{fig:kg:filter}
\end{figure}

\vspace{1pt}
{\bf Filtering of poisoning facts.} Intuitively, as they are artificially injected, poisoning facts tend to be misaligned with their neighboring entities/relations in \kgs. Thus, we propose to detect misaligned facts and filter them out to mitigate the influence of poisoning facts.
Specifically, we use \meq{eq:fit} to measure the ``fitness'' of each fact 
$v \sxrightarrow{r} v'$ and then remove $m\%$ of the facts with the lowest fitness scores.

\begin{figure*}[!ht]
    \centering
    \epsfig{file = 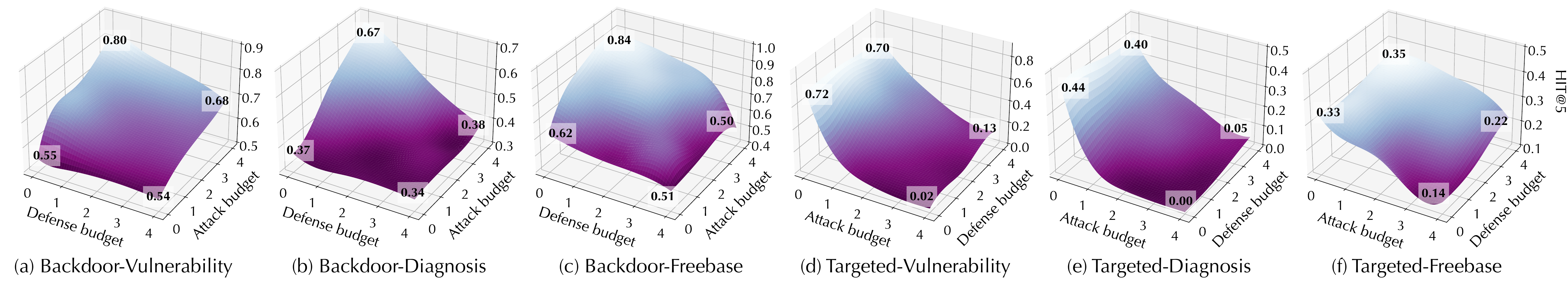, width = 165mm}
    \caption{Performance of \aco against adversarial training with respect to varying settings of attack $n_\mathrm{q}$ and defense $n_\mathrm{q}$ (note: in \zhaohan{targeted} attacks, the attack performance is measured by the \hitf drop).}
    \label{fig:cm}
\end{figure*}

Table\mref{tab:kgfilter_benign} measures the \krl performance on non-target queries $\gQ \setminus \gQ^*$ and the 
\zhaohan{Figure\mref{fig:kg:filter}} measures attack performance on target queries $\gQ^*$ as functions of $m$. We have the following observations. \mct{i} The filtering degrades the attack performance. For instance, the \hitf of \akp drops by 0.23 in the \zhaohan{backdoor} attacks against vulnerability queries as $m$ increases from 10 to 30. \mct{ii} Compared with \akp, \aco is less sensitive to filtering, which is explained by its use of both knowledge poisoning and query misguiding, with one attack vector compensating for the other. \mct{iii} The filtering also significantly impacts the \krl performance (\meg, its \hitf drops by 0.28 under $m$ = 30), suggesting the inherent trade-off between attack resilience and \krl performance. 

\begin{table}[!t]
\ting{
\renewcommand{\arraystretch}{0.9}
\centering
\setlength{\tabcolsep}{6pt}
{\footnotesize
\begin{tabular}{c|ccc}
\multirow{2}{*}{Query} & \multicolumn{3}{c}{Removal ratio ($m\%$)}\\
\cline{2-4}
& 0\% & 10\% & 30\% \\
\hline
vulnerability & 1.00 & 0.93 & 0.72 \\
diagnosis & 0.87 & 0.84 & 0.67 \\
Freebase & 0.70 & 0.66 & 0.48 \\
\hline
\end{tabular}
\caption{\zhaohan{\krl performance (\hitf) on non-target queries $\gQ \setminus \gQ^*$.} \label{tab:kgfilter_benign}}}}
\end{table}

\vspace{1pt}
{\bf Training with adversarial queries.} We further extend the adversarial training\mcite{pgd} strategy to defend against \aco. Specifically, we generate an adversarial version $q^*$ for each query $q$ using \aco and add $(q^*, \llbracket q \rrbracket)$ to the training set, where $\llbracket q \rrbracket$ is $q$'s ground-truth answer. 

We measure the performance of \aco under varying settings of $n_\text{q}$ used in \aco and that used in adversarial training, with results shown in Figure\mref{fig:cm}. 
Observe that adversarial training degrades the attack performance against the \zhaohan{backdoor} attacks (Figure\mref{fig:cm}\,a-c) especially when the defense $n_\text{q}$ is larger than the attack $n_\text{q}$. However, the defense is much less effective on the \zhaohan{targeted} attacks (Figure\mref{fig:cm}\,d-f). This can be explained by the larger attack surface of \zhaohan{targeted} attacks, which only need to force erroneous reasoning rather than \zhaohan{backdoor} reasoning. Further, it is inherently ineffective against \akp (when the attack $n_\text{q}=0$ in \aco), which does not rely on query misguiding. 

\vspace{1pt}
We can thus conclude that, to defend against the threats to \krl, it is critical to \mct{i} integrate multiple defense mechanisms and \mct{ii} balance attack resilience and \krl performance.


\subsection{Limitations}
\label{ssec:futurework}


\vspace{1pt}
{\bf Other threat models and \ting{datasets}.} While \system instantiates several attacks in the threat taxonomy in \msec{sec:threatmodel}, there are many other possible attacks against \krl. For example, if the adversary has no knowledge about the \kgs used in the \krl systems, is it possible to \ting{build surrogate KGs from scratch} or construct attacks that transfer across different \kg domains? 
\ting{Further, 
the properties of specific \kgs (\meg, size, connectivity, and skewness) may potentially bias our findings. 
}We consider exploring other threat models and datasets from other domains as our ongoing research. 

\vspace{1pt}
{\bf Alternative reasoning tasks.} We mainly focus on reasoning tasks with one target entity. There exist other reasoning tasks (\meg, path reasoning\mcite{path-reasoning} finds a logical path with given starting and end entities). Intuitively, \system is ineffective in such tasks as it requires knowledge about the logical path to perturb intermediate entities on the path. It is worth exploring the vulnerability of such alternative reasoning tasks.

\vspace{1pt}
{\bf Input-space attacks.} While \system directly operates on \kgs (or queries), there are scenarios in which \kgs (or queries) are extracted from real-world inputs. For instance, threat-hunting queries may be generated based on software testing and inspection. In such scenarios, it requires the perturbation to \kgs (or queries) to be mapped to valid inputs (\meg, functional programs). 



%% file: literature.tex
\section{Related work}


\vspace{1pt}
\indent {\bf Machine learning security.} Machine learning models are becoming the targets of various attacks\mcite{Biggio:2018:pr}: adversarial evasion crafts adversarial inputs to deceive target models\mcite{goodfellow:fsgm,carlini-attack}; model poisoning modifies target models' behavior by polluting training data\mcite{model-reuse}; backdoor injection creates trojan models such that trigger-embedded inputs are misclassified\mcite{trojannn, changjiang-ssl}; functionality stealing constructs replicate models functionally similar to victim models\mcite{model-stealing}. In response, intensive research is conducted on improving the attack resilience of machine learning models. For instance,  existing work explores new training strategies (\meg, adversarial training)\mcite{pgd} and detection mechanisms\mcite{Gehr:2018:sp, changjiang-tdsc} against adversarial evasion. Yet, such defenses often fail when facing adaptive attacks\mcite{Athalye:2018:icml,deepsec}, resulting in a constant arms race.

\vspace{1pt}
{\bf Graph learning security.} Besides general machine learning security, one line of work focuses on the vulnerability of graph learning\mcite{gcn, gat, Kaidi:ijcai:2019}, including adversarial\mcite{Zugner:kdd:2018, Binghui:ccs:2019, poison-embedding:icml:2019}, poisoning\mcite{graph-poisoning}, and backdoor\mcite{graphbackdoor} attacks. \ting{This work differs from existing attacks against graph learning in several major aspects. \mct{i} Data complexity -- while \kgs are special forms of graphs, they contain much richer relational information beyond topological structures. \mct{ii} Attack objectives -- we focus on attacking the logical reasoning task, whereas most existing attacks aim at the classification\mcite{Zugner:kdd:2018, Binghui:ccs:2019, graph-poisoning} or link prediction task\mcite{poison-embedding:icml:2019}.
\mct{iii} Roles of graphs/KGs -- we target \kgr systems with \kgs as backend knowledge bases while existing attacks assume graphs as input data to graph learning.
\mct{iv} Attack vectors -- we generate plausible poisoning facts or bait evidence, which are specifically applicable to \kgr; in contrast, previous attacks directly perturb graph structures\mcite{Binghui:ccs:2019, poison-embedding:icml:2019, graph-poisoning} or node features\mcite{Zugner:kdd:2018, graphbackdoor}.}




\vspace{1pt}
{\bf Knowledge graph security.} 
The security risks of \kgs are gaining growing attention\mcite{kg-poisoning-ijcai, kg-poisoning-acl, kge-attack-emnlp,kg-link-attack,kg-deceive-iclr}. Yet, most existing work focuses on the task of link prediction (\kg completion) and the attack vector of directly modifying \kgs. This work departs from prior work in major aspects: 
\mct{i} 
we consider reasoning tasks (\meg, processing logical queries), which require vastly different processing from predictive tasks (details in Section \msec{sec:background});
\mct{ii} existing attacks rely on directly modifying the topological structures of KGs (\meg, adding/deleting edges) without accounting for their semantics, while we assume the adversary influences \krl through indirect means with semantic constraints (\meg, injecting probable relations or showing misleading evidence);
\mct{iii} we evaluate the attacks in real-world \krl applications; and \mct{iv} we explore potential countermeasures against the proposed attacks.

%% file: conclusion.tex
\section{Conclusion}

This work represents a systematic study of the security risks of knowledge graph reasoning (\krl). We present \system, a new class of attacks that instantiate a variety of threats to \krl. We demonstrate the practicality of \system in domain-specific and general \krl applications, raising concerns about the current practice of training and operating \krl. We also discuss potential mitigation against \system, which sheds light on applying \krl in a more secure manner. 


%% file: appendix.tex
\appendix





\section{Notations}
\label{sec:notation}

Table\mref{tab:notations} summarizes notations and definitions used through this paper.

\begin{table}[!ht]{
\centering
\small
\renewcommand{\arraystretch}{1}
\begin{tabular}{c | l }

Notation & Definition \\
\hline
\hline
\multicolumn{2}{l}{{Knowledge graph related}} \\
\hline
$\gG$ & a knowledge graph (\kg) \\
$\gG'$ & a surrogate knowledge graph\\
$\langle v, r, v' \rangle$ & a \kg fact from entity $v$ to $v'$ with relation $r$ \\
$\gN, \gE, \gR$ & entity, edge, and relation set of $\gG$\\
$\gG^+$ & the poisoning facts on \kg \\

\hline
\multicolumn{2}{l}{{Query related}} \\
\hline
$q$ & a single query \\
$\llbracket q \rrbracket$ & $q$'s ground-truth answer(s) \\
$a^*$ & the targeted answer \\
$\gA_q$ & anchor entities of query $q$ \\
$p^*$ & the trigger pattern \\
$\gQ$ & a query set \\
$\gQ^*$ & a query set of interest (each $q\in\gQ^*$ contains $p^*$)\\
$q^+$ & the generated bait evidence\\
$q^*$ & the infected query, \mie $q^* = q \wedge q^+$ \\
\hline
\multicolumn{2}{l}{{Model or embedding related}} \\
\hline
$\phi$ & a general symbol to represent embeddings \\
$\phi_{\gG}$ & embeddings of all \kg entities \\
$\phi_v$ & entity $v$'s embedding  \\
$\phi_{q}$ & $q$'s embedding \\
$\phi_{\gG^+}$ & embeddings we aim to perturb \\
$\phi_{q^+}$ & $q^+$'s embedding \\
$\psi$ & the logical operator(s) \\
$\psi_r$ & the relation ($r$)-specific operator \\
$\psi_\wedge$ & the intersection operator  \\
\hline
\multicolumn{2}{l}{{Other parameters}} \\
\hline
$n_\mathrm{g}$ & knowledge poisoning budget \\
$n_\mathrm{q}$ & query misguiding budget \\
\hline
\end{tabular}
\caption{Notations, definitions, and categories. \label{tab:notations}}}
\end{table}

\section{Additonal details}
\label{sec:comp_info}

\subsection{\zhaohan{\kgr training}}

Following\mcite{query2box}, we train \kgr in an end-to-end manner. Specifically, given \kg $\gG$ and the randomly initialized embedding function $\phi$ and transformation function $\psi$, we sample a set of query-answer pairs $(q, \llbracket q \rrbracket)$ from $\gG$ to form the training set and optimize $\phi$ and $\psi$ to minimize the loss function, which is defined as the embedding distance between the prediction regarding each $q$ and $\llbracket q \rrbracket$.

\subsection{Parameter setting} 
\label{ssec:setting}

Table\mref{tab:param} lists the default parameter setting used in \msec{sec:expt}.

\begin{table}[!ht]{\footnotesize
\centering
\setlength{\tabcolsep}{4pt}
\renewcommand{\arraystretch}{1}
\begin{tabular}{c|r|l}
 Type & Parameter & Setting \\
\hline
\hline 
\multirow{6}{*}{\krl} & $\phi$ dimension & 300 \\
 & $\phi$ dimension (surrogate) & 200\\
 & $\psi_r$ architecture & 4-layer FC\\
 & $\psi_\wedge$ architecture & 4-layer FC\\
 & $\psi_r$ architecture (surrogate) & 2-layer FC\\
 & $\psi_\wedge$ architecture (surrogate) & 2-layer FC\\
\hline
\multirow{5}{*}{Training} & Learning rate & 0.001 \\
& Batch size & 512 \\
& \krl epochs & 50000 \\
& \system optimization epochs & 10000 \\
& Optimizer (\krl and \system) & Adam \\
\hline
\multirow{3}{*}{Other} & $n_\mathrm{g}$ & 100 \\
& $n_\mathrm{q}$ & 2 \\
\hline
\end{tabular}
\caption{Default parameter setting.\label{tab:param}}}
\end{table}

\subsection{Extension to \zhaohan{targeted} attacks}
\label{sec:extension}

It is straightforward to extend \system to \zhaohan{targeted} attacks, in which the adversary aims to simply force \krl to make erroneous reasoning over the target queries $\gQ^*$. To this end, we may maximize the distance between the embedding $\phi_q$ of each query $q \in \gQ^*$ and its ground-truth answer $\llbracket q \rrbracket$. 

Specifically, in knowledge poisoning, we re-define the loss function in \meq{eq:kp} as:
\begin{equation}
\label{eq:kp:untar}
\begin{split}
\ell_\text{kp}(\phi_{\gG^+}) = \, & \mathbb{E}_{q \in \gQ \setminus \gQ^*} \Delta (\psi(q ; \phi_{\gG^+}), \phi_{\llbracket q \rrbracket}) - \\
& \lambda \mathbb{E}_{q \in \gQ^*} \Delta (\psi(q ; \phi_{\gG^+}), \phi_{\llbracket q \rrbracket})
\end{split}
\end{equation}
In query misguiding, we re-define \meq{eq:qp} as:
\begin{equation}
\label{eq:qp:untar}
\ell_\text{qm}(\phi_{q^+})
 =  - \Delta (\psi_{\wedge}(\phi_q, \phi_{q^+}),\, \phi_{\llbracket q \rrbracket})
\end{equation}
The remaining steps are the same as the \zhaohan{backdoor} attacks.

\subsection{Additional results}
\label{ssec:add-results}

This part shows the additional experiments as the complement of section\msec{sec:expt}.

\begin{figure*}[!ht]
    \centering
    \epsfig{file = 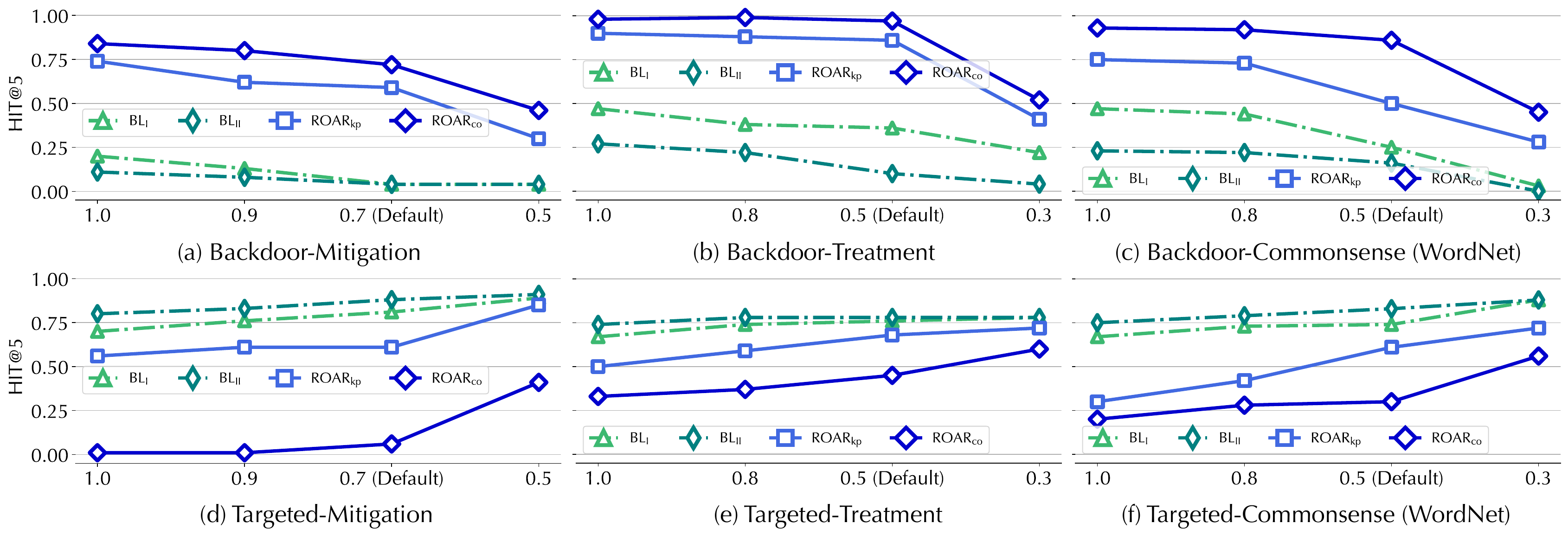, width = 155mm}
    \caption{\akp and \aco performance with varying overlapping ratios between the surrogate and target \kgs, measured by \hitf after the attacks \zhaohan{on other query tasks besides Figure\mref{fig:surkg}}.}
    \label{fig:surkg2}
\end{figure*}

\zhaohan{{\bf Additional query tasks under variant surrogate \kgs.} Figure\mref{fig:surkg2} presents the attack performance on other query tasks that are not included in Figure\mref{fig:surkg}. We can observe a similar trend as concluded in\msec{ssec:factor}.}

\begin{figure*}[!ht]
    \centering
    \epsfig{file = 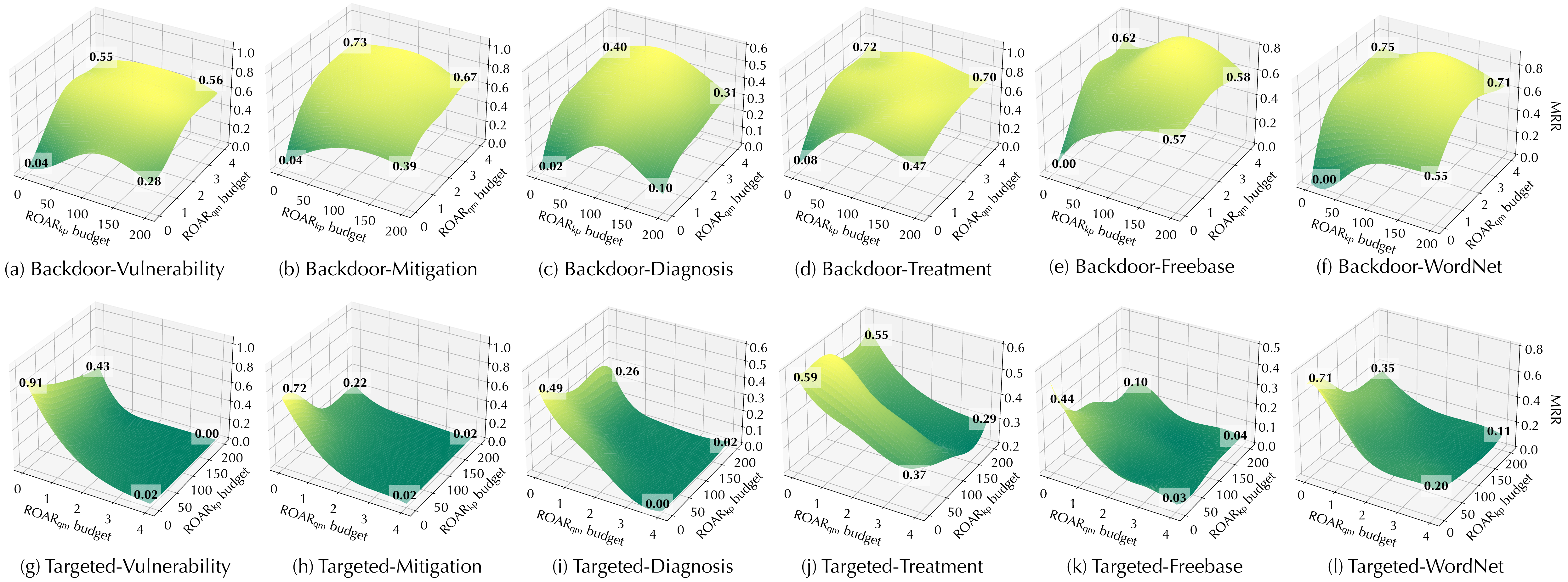, width = 162mm}
    \caption{\aco performance with varying budgets (\akp -- $n_\mathrm{g}$, \aqp -- $n_\mathrm{q}$). The measures are the absolute \mrr after the attacks.}
    \label{fig:budget:mrr}
\end{figure*}

\begin{figure}[!tp]
    \centering
    \epsfig{file = 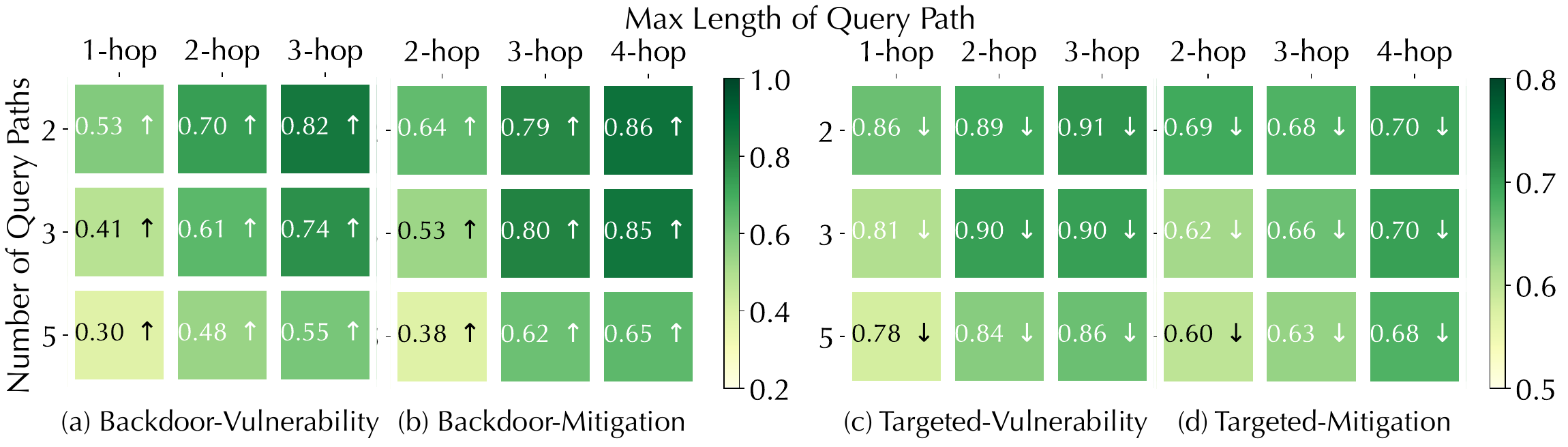, width = 88mm}
    \caption{\aco performance (\mrr) under different query structures in Figure\mref{fig:qstruc}, indicated by the change ($\uparrow$ or $\downarrow$) before and after the attacks.}
    \label{fig:diff:struc:mrr}
\end{figure}

\zhaohan{{\bf MRR results}}. Figure\mref{fig:diff:struc:mrr} shows the \mrr of \aco with respect to different query structures, with observations similar to Figure\mref{fig:diff:struc}. 
Figure\mref{fig:budget:mrr} shows the \mrr of \system with respect to attack budgets ($n_\mathrm{g}$, $n_\mathrm{q}$), with observations similar to Figure\mref{fig:budget:hit}.

